\documentclass{article}
\usepackage[top=1.25in,bottom=1in,left=1in,right=1in]{geometry}
\usepackage[utf8]{inputenc} 
\usepackage[english]{babel}

\usepackage{amsmath}
\usepackage{graphicx}
\usepackage[colorlinks=true, allcolors=blue]{hyperref}
\usepackage{natbib}
\usepackage{multicol}
\usepackage{multirow}
\usepackage{adjustbox}
\usepackage{booktabs}
\usepackage{subcaption}
\usepackage{listings}
\usepackage{tabularx}
\usepackage{longtable}

\usepackage{pgfplots}
\usepackage{floatrow}
\usepackage{listings}

\usepackage[utf8]{inputenc}
\usepackage{etoolbox}

\usepackage{authblk}
\makeatletter
\renewcommand{\@makefnmark}{}  
\makeatother

\title{Novelty and Impact of Economics Papers}
\author{Chaofeng Wu\thanks{Department of Computer Science, Northwestern University}}

\begin{document}
\maketitle
\footnotetext{We thank Annie Liang, Ben Golub, Jeff Ely, Jessica Hullman, and Anqi Zhang for helpful comments. We also thank the Quest high performance computing facility at Northwestern University for computational resources and staff assistance.}
\pagenumbering{gobble}
\begin{abstract}

We propose a framework that recasts scientific novelty not as a single attribute of a paper, but as a reflection of its position within the evolving intellectual landscape. We decompose this position into two orthogonal dimensions: \textit{spatial novelty}, which measures a paper's intellectual distinctiveness from its neighbors, and \textit{temporal novelty}, which captures its engagement with a dynamic research frontier. To operationalize these concepts, we leverage Large Language Models to develop semantic isolation metrics that quantify a paper's location relative to the full-text literature.
Applying this framework to a large corpus of economics articles, we uncover a fundamental trade-off: these two dimensions predict systematically different outcomes. Temporal novelty primarily predicts citation counts, whereas spatial novelty predicts disruptive impact.
This distinction allows us to construct a typology of semantic neighborhoods, identifying four archetypes associated with distinct and predictable impact profiles. Our findings demonstrate that novelty can be understood as a multidimensional construct whose different forms, reflecting a paper's strategic location, have measurable and fundamentally distinct consequences for scientific progress.

\end{abstract}
\newpage \pagenumbering{arabic} \setcounter{page}{1}
\section{Introduction}

A central challenge for researchers, funding agencies, and academic institutions is to identify which new ideas are most likely to generate significant scientific impact. While novelty is widely regarded as a key driver of impact, its complex nature and the mechanisms through which it operates remain poorly understood. This challenge reflects a fundamental tension in research strategy: should efforts focus on work with high uniqueness, characterized by profound originality in questions and methods, or on work with good timing that taps into "hot" topics currently capturing the scientific community's attention? 
Although both uniqueness and timing are recognized as critical dimensions of novelty, existing measures typically fail to disentangle them.
Citation-based approaches \citep{uzzi2013atypical,lee2015creativity,trapido2015novelty} overlook a paper's semantic content, while keyword-based methods \citep{azoulay2011incentives,mishra2016quantifying,arts2025beyond} cannot quantify its departure from prior work.

In this study, we introduce a new conceptual framework that recasts novelty not as an intrinsic property, but as a positional good, defined by a paper's location within the evolving intellectual landscape. We characterize this position along two orthogonal dimensions. 
\textbf{Spatial novelty} measures a paper's intellectual distinctiveness by quantifying its semantic distance from all prior work at the time of publication, capturing the uniqueness of the work. \textbf{Temporal novelty}, in contrast, measures a paper's engagement with a dynamic and evolving research frontier by quantifying the rate of change in its immediate semantic neighborhood, capturing the timing of the work.

To operationalize this framework, we introduce a set of semantic isolation metrics that directly quantify how far a paper's semantic content deviates from other work, a measurement that aligns with the fundamental definition of novelty. 
Leveraging recent advances in Natural Language Processing (NLP) and Large Language Models (LLMs), we develop a pipeline to analyze the full text of research papers and compute these metrics based on semantic embeddings. 
Using a custom dataset containing full text of economic papers, we validate these metrics by demonstrating their strong predictive power for two distinct measures of impact: citation counts and disruption.

Building on these validated metrics, we construct composite scores for spatial novelty and temporal novelty, uncovering a critical trade-off in research strategy. Our analysis yields three main discoveries. 
First, temporal novelty is the primary predictor of citations, whereas spatial novelty is the main predictor of a paper's disruptive influence-its tendency to render earlier work obsolete rather than build upon it. 
Second, this distinction allows for a more refined classification of intellectual neighborhoods into four archetypes: ``Consolidating'', ``Outlying'', ``Trendy'', and ``Trailblazing'', each with a distinct and predictable impact profile. 
Finally, we find that the type of papers located in the Trailblazing neighborhoods --- who successfully balance high spatial and high temporal novelty --- have a disproportionately high probability of being impactful on both dimensions, suggesting a synergistic interaction between intellectual distinctiveness and strategic timing. Our findings demonstrates that novelty is multidimensional, with measurable and systematically different consequences for scientific progress.

The remainder of the introduction elaborates on each component of our contribution. We first detail our proposed conceptual framework. We then describe the construction and validation of our semantic isolation metrics. Next, we provide a more detailed overview of our empirical findings on the relationship between the novelty typology and scientific impact. The section concludes with a discussion of the related literature.

\paragraph{Proposed Framework}
We first establish a framework for measuring the multidimensional nature of both novelty and impact. Following \cite{kuhn1970structure}, who argued that scientific ideas are embedded in the text of the scientific literature, we propose that novelty can be measured by quantifying a paper's semantic neighborhood. We decompose this concept into two dimensions. \textbf{Spatial novelty} measures the semantic isolation of a paper from all prior work at its publication date, capturing its intellectual distinctiveness. \textbf{Temporal novelty} measures the rate of semantic change within a paper's immediate intellectual neighborhood, reflecting its engagement with a dynamic research frontier.
The two dimensions describe a paper's position within the evolving intellectual landscape.

To measure impact, we employ two widely used but conceptually distinct metrics. The first is citation count, which reflects a paper's popularity and the research community's collective judgment of its value. The second is the disruption index \citep{wu2019large}, which measures whether a paper's contribution renders prior work obsolete (disruptive) or serves as a cumulative link between past and future research (consolidating).

\paragraph{Semantic Isolation Metrics and Validation}
To quantify spatial and temporal novelty, we design a set of semantic isolation metrics based on papers' positions in a high-dimensional semantic space constructed from text embeddings. 
In this space, greater semantic isolation, measured as the distance from contemporary works, is our operational definition of novelty, capturing a paper's intellectual distinctiveness.

Our metrics include both point-in-time and dynamic measures. \textbf{Point-in-time metrics} quantify how semantically distant a paper is from all existing work at a specific moment. To capture this, we measure several features of a paper's local neighborhood, including the number of neighbors within a fixed radius, the distance to its $k-$nearest neighbors, the average distance to its k$-$nearest neighbors, and the local density estimated using a Gaussian kernel. This set of metrics captures not only papers that are uniformly distant from all prior work but also those that are close to a small cluster while being far from the vast majority of other papers.

While point-in-time metrics assess a paper's isolation at publication, \textbf{dynamic metrics} capture how its semantic neighborhood evolves over time. A rapid change in a paper's neighborhood - indicated by many new papers emerging nearby in the recent past or near future - suggests it is situated within a research neighborhood with strong momentum. Conversely, a stable neighborhood suggests the paper occupies a more settled intellectual terrain. These dynamic metrics can be constructed retrospectively (based on past changes) or prospectively (based on future changes), and we only use retrospective version in our main results.

Using a custom dataset containing 105,082 full text of economic papers, 
we validate our semantic isolation metrics by demonstrating that they are strong predictors of both citations and disruption, providing information that is complementary to existing measures. 
Using the paper's field of study as a baseline, we show that our isolation metrics significantly improve predictive performance and contribute to the prediction even when a broad set of features exist.

\paragraph{Typology of Novelty and Relation to Impact}
Having validated the semantic isolation metrics, we employ them to construct our two-dimensional novelty framework. We create a composite \textbf{spatial novelty score} by applying Principal Component Analysis (PCA) to the point-in-time isolation metrics. Similarly, we construct a \textbf{temporal novelty score} using PCA on the retrospective dynamic metrics. By using only contemporaneous and retrospective information, our framework ensures that a paper's novelty is measured using only information available at the time of its publication.

The resulting spatial and temporal novelty scores exhibit a negative correlation, revealing a structural tension between them. 
This suggests that it is empirically challenging for a single work to be simultaneously highly distinctive (high spatial novelty) and perfectly timed within an emerging research wave (high temporal novelty).

This two-dimensional space allows for a typology of a paper's position in the intellectual landscape. By defining papers above the median on each dimension as "high", we classify a paper's intellectual neighborhoods into four distinct profiles: Consolidating (low spatial, low temporal), Outlying (high spatial, low temporal), Trendy (low spatial, high temporal), and Trailblazing (high spatial, high temporal). Each archetype represents a unique type of position of a paper's position in the evolving intellectual landscape.

Applying this framework, we analyze the relationship between novelty and impact. We find that temporal novelty is the primary predictor of high citations; papers with high temporal novelty (in Trendy or Trailblazing neighborhoods) are significantly more likely to become highly cited papers, while spatial novelty provides only modest gains. Conversely, spatial novelty predicts disruption; papers with high spatial novelty (in Outlying or Trailblazing neighborhoods) are far more likely to be disruptive, whereas temporal novelty has a slight negative association with disruption. Finally, papers in Trailblazing neighborhoods, who successfully navigate the trade-off between the two dimensions, have a significantly higher probability of being both highly cited and highly disruptive, suggesting a synergistic interaction that amplifies a paper's transformative potential.

\subsection{Related Work}

\paragraph{Measures of novelty.}
The measurement of scientific novelty has traditionally relied on citation networks and textual data \citep{zhao2025review}. A prominent citation-based approach is the atypicality score, which quantifies novelty based on unconventional combinations of referenced journals \citep{uzzi2013atypical}. This concept has been extended to account for same-year publications \citep{lee2015creativity} and to measure the unusualness of knowledge recombination within a specific field \citep{trapido2015novelty}.

Textual data offers a more direct window into a paper's content. One common approach assumes that new words or word combinations signal novel ideas. For example, novelty has been measured using the average age of MeSH\footnote{MeSH (Medical Subject Headings) is the NLM controlled vocabulary thesaurus used for indexing articles in PubMed.} terms \citep{azoulay2011incentives}, their age and reuse frequency \citep{mishra2016quantifying}, and the tracking of new word combinations and their subsequent reuse \citep{arts2025beyond}. However, keyword-based methods can be biased by polysemy and often overlook deeper semantic relationships. To mitigate this, other work has extracted entities as proxies for knowledge \citep{luo2022combination,chen2024exploring} or identified novel contribution sentences using semantic analysis \citep{wang2023measuring}. Another approach assessed novelty by computing semantic distances between a paper's references, represented by title embeddings \citep{shibayama2021measuring}.
Our method departs from these proxies by analyzing the full text of a paper and using LLMs to generate comprehensive semantic representations. This approach captures a richer representation of a paper's content than methods relying solely on titles, abstracts, or keywords. By computing semantic distances between papers, we construct metrics that directly measure a paper's departure from the existing literature and its engagement in research frontier.

\paragraph{Validation of novelty measures.}
Novelty measures are typically validated through two primary approaches. Direct validation involves correlating proposed metrics with ground-truth labels derived from expert evaluations or peer review outcomes \citep{tahamtan2018creativity, shibayama2021measuring}. Indirect validation, which is more common, assesses a metric's correlation with established novelty indices or its predictive power for outcomes like citation counts \citep{luo2022combination, foster2021surprise, shi2023surprising, arts2025beyond}. We adopt the latter approach, validating our measures by evaluating their ability to predict citation counts and disruption.

\paragraph{Conceptualization of novelty.}
Scientific novelty has been conceptualized through various theoretical lenses \citep{zhao2025review}. A foundational dichotomy distinguishes between the introduction of entirely new elements \citep{berlyne1960conflict, yan2020impact, luo2022combination} and the novel recombination of existing elements, a concept rooted in Schumpeterian innovation theory \citep{schumpeter1983theory, nelson1985evolutionary, guetzkow2004originality}. Other frameworks conceptualize novelty in terms of network structures, where new connections bridge previously disconnected knowledge communities \citep{foster2015tradition, hofstra2020diversity}, or from a psychological perspective, where novelty is a function of surprise relative to a reader's cognitive schema \citep{foster2021surprise}. 
We use a paper's position within the intellectual landscape as its novelty. This captures not only the semantic difference between the focal paper and other work, but also the evolvement of the focal paper's semantic neighborhood.

\paragraph{Typologies of Novelty.}
Prior literature has also sought to create typologies of novelty, generally along two distinct axes. The first stream differentiates novelty by its form or mechanism. This includes distinctions between novelty arising from the age and volume of concepts \citep{mishra2016quantifying}, the recombination of existing knowledge versus the creation of new elements \citep{yan2020impact}, or between a paper's content and its context \citep{shi2023surprising}. A second stream distinguishes novelty by its locus or domain of contribution, such as new theory, new methodology, or new empirical results \citep{leahey2023types}.
Our framework contributes to this literature by proposing a new form-based typology---spatial versus temporal novelty---that is conceptually orthogonal to the locus of contribution. A key advantage of our measurement approach is its ability to operationalize this orthogonality. By decomposing each paper into distinct insight types, we can measure the spatial and temporal novelty of a paper's theory, its methodology, or its core question-conclusion pair, thereby providing a more granular understanding of scientific innovation.

\paragraph{NLP and LLMs in text analysis.}
Computational analysis of scientific texts has evolved significantly. Early work relied on sparse vector representations such as term frequency–inverse document frequency (TF–IDF) and probabilistic topic models like Latent Dirichlet Allocation (LDA) \citep{blei2003latent} to map research trends \citep{gupta2022prediction, egger2022topic}. More recent methods leverage dense vector representations (embeddings) to model document semantics more effectively. Embedding-based models like SPECTER \citep{cohan2020specter} and domain-specific word embeddings \citep{tshitoyan2019unsupervised} have improved performance on tasks such as classification, clustering, and citation prediction by capturing nuanced contextual relationships.

Building on these foundations, large language models (LLMs) now enable direct synthesis and representation of scientific text at an unprecedented scale. Current applications include screening for systematic reviews \citep{dennstadt2024title}, generating retrieval-augmented literature surveys \citep{han2024automating}, classifying research articles \citep{raja2023using}, and predicting bibliometric outcomes \citep{vital2025predicting}. LLMs have also proven effective in summarization tasks, producing outputs for both expert and lay audiences \citep{pakull2024wispermed, ding2024ccsbench} and even drafting review articles \citep{wang2024using}. Our work contributes to this literature by using LLMs not just for summarization but to create the rich semantic representations that underpin our measurement of novelty.

\section{Framework}
\label{sec:framework}

This section develops the conceptual and measurement framework for our analysis. We first define our two measures of scientific impact---citation count and the disruption index---which capture distinct dimensions of a paper's influence. 
We then reconceptualize novelty in positional terms, introducing spatial and temporal novelty to capture a paper's location in the intellectual landscape.
Finally, we illustrate the semantic isolation metrics used to operationalize our framework.

\subsection{Measuring Multidimensional Impact}
To test our hypothesis that different forms of novelty predict different types of scientific influence, we must first define impact in a multidimensional manner. A paper might attract widespread attention, reflected in high citation counts, while another might receive modest immediate attention yet fundamentally redirect its field's trajectory. To capture these distinct pathways, we employ two complementary metrics that capture distinct dimensions of scientific impact: citation count and the disruption index.

\paragraph{Citation Count: Measuring Scholarly Attention.}
Our first measure, cumulative citation count, serves as a standard proxy for scholarly attention and influence. Citations reflect the research community's collective assessment of a paper's value and its direct utility in subsequent scholarship. 

\paragraph{Disruption Index: Measuring Field Redirection.}
Our second measure captures a paper's capacity to redirect a field's intellectual trajectory. 
We employ the Disruption Index (DI) developed by \cite{funk2017dynamic} and \cite{wu2019large}, which we classify as an impact measure rather than a novelty measure \cite{Leibel2024disruption}.
Whereas citations measure the magnitude of influence, the DI measures its nature: whether a paper's contribution displaces prior work or builds cumulatively upon it. The index is derived from the citation network surrounding a focal paper (FP) and is defined as:
\[
DI = \frac{N_F-N_B}{N_F+N_B+N_R}
\]
where $N_F$ represents papers citing only the focal paper, $N_B$ represents papers citing both the focal paper and its references, and $N_R$ serves as a normalizing term for papers citing the focal paper's references but not the focal paper itself. 

A high positive score indicates a "disruptive" paper, as subsequent work cites it as a new origin point, displacing its intellectual predecessors. A negative score suggests a "consolidating" paper that integrates prior knowledge, prompting subsequent work to cite both the focal paper and its antecedents. 
This dual-measurement strategy allows us to test whether different dimensions of novelty systematically produce different forms of impact.

\subsection{Novelty as Semantic Position}
The study of scientific progress hinges on the concept of novelty, yet its measurement remains a central challenge. Prevailing methods, such as the widely-used atypicality score that measures unconventional reference combinations \citep{uzzi2013atypical}, have provided powerful insights. However, these approaches share a common conceptual foundation: they treat novelty as an intrinsic attribute of a paper, inferred from its inputs (e.g., references) or its internal content (e.g., new keywords). This view, while intuitive, limits our ability to understand the strategic and ecological dynamics of scientific discovery.


We propose a fundamental shift in perspective, reconceptualizing novelty not as an intrinsic property but as a relational construct. 
In our view, a paper's novelty is a function of its location within a high-dimensional semantic space, characterized by both its static isolation and its dynamic evolution.


To characterize a paper's position, we decompose it into two orthogonal dimensions. These dimensions describe its relationship with the dynamic ``semantic neighborhood'' it inhabits:

\begin{itemize}
\item \textbf{Spatial novelty} measures a paper's static position in the semantic space. It quantifies intellectual distinctiveness by measuring a paper's isolation from the accumulated stock of prior knowledge.
\item \textbf{Temporal novelty} measures the dynamics of its position. It captures strategic timing by measuring the change in a paper's immediate semantic over the period leading up to publication, reflecting its engagement with an evolving research frontier.
\end{itemize}

These two dimensions are central to our analysis, as we hypothesize they correspond to distinct strategic positions with fundamentally different consequences for scientific impact. To operationalize this framework, we develop a family of semantic isolation metrics, which we describe next. These metrics are designed to quantify a paper's static isolation (for spatial novelty) and the change in its isolation over time (for temporal novelty) using text embeddings.

\subsection{Semantic Isolation Metrics}

We construct semantic isolation metrics to quantify both spatial and temporal novelty. Each metric is based on the semantic distance between papers, computed by embedding every paper's full text into a high-dimensional vector space (see Section \ref{sec:pipeline_for_embedding}) and measuring pairwise distances.
For spatial novelty, we define four point-in-time metrics, each capturing a distinct facet of a paper's semantic distinctiveness relative to the existing literature. For temporal novelty, we derive dynamic metrics by tracking how each point-in-time measure evolves across successive time windows around a paper's publication (e.g., three years pre-publication versus publication). The resulting slopes serve as our indicators of well-timed engagement with emerging research frontiers.
We synthesize these individual measures into two composite scores, as detailed in Section \ref{sec:construct_novelty_score}.

In this section, we provide a conceptual illustration of semantic isolation metrics. A full technical description of these metrics appears in Appendix \ref{app-sec:isolation_indicators}.

\begin{figure}[h!]
    \centering
    \includegraphics[width=0.5\linewidth]{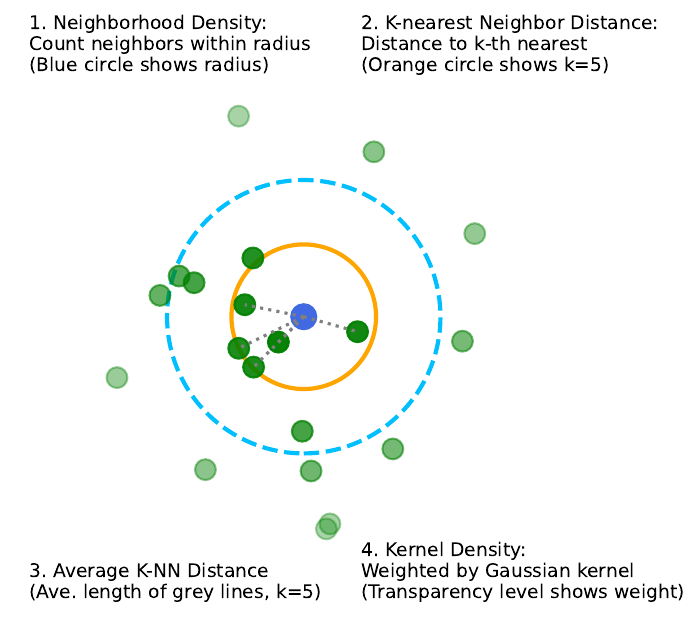}
    \caption{Illustration of point-in-time isolation metrics. The blue point is the focal paper, and the green points are other papers.}
    \label{fig:point_in_time_metrics}
\end{figure}

\paragraph{Point-in-Time Isolation Metrics.}
    Point-in-time metrics assess a paper's semantic isolation at a given moment. As illustrated in Figure \ref{fig:point_in_time_metrics}, for each focal paper (blue point), we consider its relationship to other work (green points), as measured by four complementary metrics. 

\begin{enumerate}
\item \textbf{Neighborhood Count:} The number of neighboring papers within a given radius. 
\item \textbf{$k-$Nearest Neighbor ($k$-NN) Distance:} The distance to the $k$-th nearest neighbor. A larger distance implies that a paper must "reach" further into the semantic space to find even its closest peers.
\item \textbf{Average $k$-NN Distance:} The average distance to the $k$ nearest neighbors. This provides a more robust measure of local isolation than the single $k$-NN distance, as it is less sensitive to the position of a single neighbor.
\item \textbf{Kernel Density Estimate:} A smooth, weighted measure of local density computed using a Gaussian kernel. This method provides a more continuous measure of isolation than discrete neighbor counts by down-weighting more distant papers.
\end{enumerate}
These complementary metrics provide a complete picture of a paper's semantic neighborhood. 
This approach allows us to distinguish whether a paper is truly isolated from all prior work, or if it belongs to a small, nascent cluster that is itself distant from the dense core of the literature.
Collectively, these metrics provide a rich characterization of a paper's spatial novelty. 

\paragraph{Dynamic Isolation Metrics.}
While point-in-time metrics capture static isolation, our dynamic metrics measure the evolution of a paper's semantic neighborhood. These metrics are designed to quantify temporal novelty by assessing whether a paper is situated in a stable intellectual area or a rapidly evolving research frontier. 
We construct these by calculating a point-in-time isolation metric, $M(t)$, at different time horizons relative to the focal paper's publication date ($t=0$), as shown in Figure \ref{fig:dynamic_metrics}. We measure change from two perspectives:

\begin{figure}[h!]
    \centering
    \includegraphics[width=0.9\linewidth]{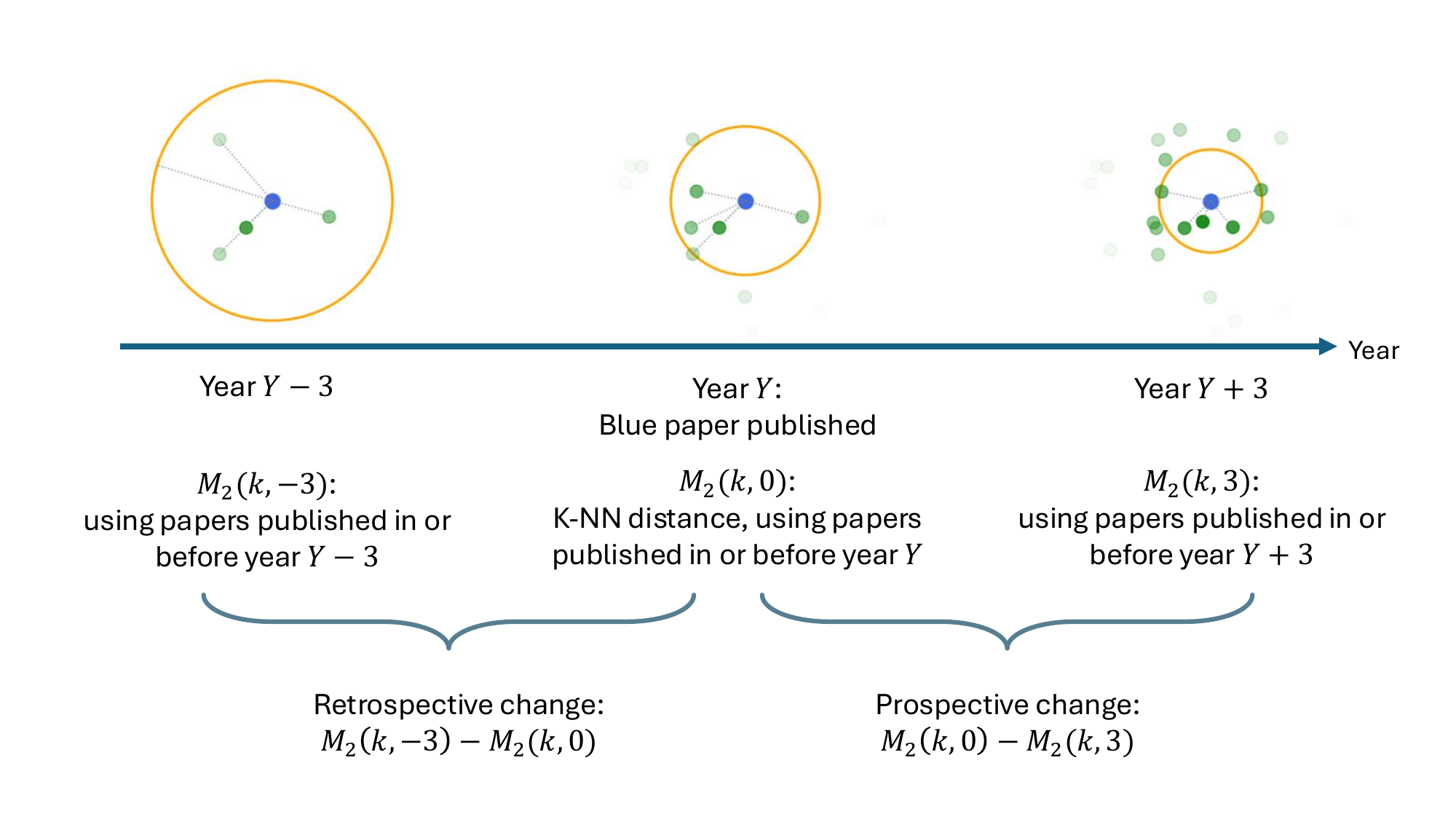}
    \caption{Illustration of dynamic isolation metrics. Given the focal paper published in year $Y$, $M_2(k, t)$ represents the $k-$NN distance using papers that published in or before year $Y+t$. For example, $M_2(k, -3)$ is calculated with papers that published in or before year $Y-3$.}
    \label{fig:dynamic_metrics}
\end{figure}

\begin{itemize}
    \item \textbf{Retrospective Change}: 
    We compute the change in an isolation metric over the period leading up to the focal paper's publication (e.g., from $t=-3$ to $t=0$). For instance, the change in $k$-NN distance, $M_2(k, -3) - M_2(k, 0)$, measures how much denser the neighborhood became in the three years prior to publication. A large positive change indicates that the paper is entering a rapidly solidifying, "hot" research frontier. This is our primary measure for constructing the temporal novelty score, as it relies only on information available at publication.
    \item \textbf{Prospective Change:} We also compute the change in the years following publication (e.g., from $t=0$ to $t=3$). The difference $M_2(k, 0)-M_2(k, 3)$ measures how the neighborhood evolves after publication of the focal paper. 
    Because perspective change relies on future information unavailable at the time of publication, we exclude it from our main results and instead use it to validate our temporal novelty concept (see Appendix \ref{app-sec:temporal_isolation}).
\end{itemize}

For example, a paper on behavioral economics published in 2005 exhibits high temporal novelty if many similar papers appeared between 2002 and 2004 (high retrospective change), indicating it is well-timed to an emerging frontier, even if its spatial novelty is modest.

In addition to these two main types of semantic isolation metrics, we also include baseline paper-level mean distance, which captures basic features of semantic distance. More details are in Appendix \ref{app-sec:baseline_semantic_metric}.

\section{Data}
\label{sec:data}

Our empirical analysis relies on a large corpus of full-text economics articles processed through a custom natural language understanding pipeline. In this section, we (1) describe our data sources, (2) explain how we constructed the analysis sample, and (3) summarize the steps used to generate our semantic isolation metrics. A key contribution of this paper is the creation of a novel, integrated dataset containing papers' full text and their semantic embeddings. We also contribute a reproducible pipeline for extracting these embeddings from article full text.

\subsection{Data Source and Dataset Construction}

Our analysis draws on a comprehensive corpus of scholarly articles in economics. 
The initial dataset comprises 105,082 articles published between 1980 and 2020, sourced from RePEc, Google Scholar, and SciSciNet \citep{lin2023sciscinet}. From SciSciNet, we obtain key bibliometric variables, including the disruption index and the atypicality score. We supplement this with citation counts collected from Google Scholar in December 2024.

To ensure a robust measurement of novelty and impact, we partition our sample by time. Papers published between 1980 and 1995 form a "knowledge base," providing the necessary historical context to measure the novelty of subsequent work. Our analysis focuses on a "focal period" cohort of papers published between 1996 and 2015. This design allows us to observe impact outcomes over a meaningful time horizon (at least 9 years post-publication) while grounding our novelty measures in a deep historical context. After filtering for data quality and completeness, our final analysis dataset contains 52,766 papers.

Table \ref{tab:data_description} provides descriptive statistics for the focal sample, disaggregated by subfield (classified using a large language model). The data reveal substantial heterogeneity across fields in both citation patterns and the prevalence of disruptive work. For instance, Development and Agricultural Economics have the highest median citation counts and disruptive rates, respectively, while Microeconomic Theory exhibits lower values on both dimensions. As expected, the distribution of citations is highly right-skewed across all fields.

\begin{table}[h!]
    \centering
\begin{tabular}{lrllll}
\toprule
Field & Count & Mean cite (std) & Median cite & 75-th cite & Disruptive rate \\
\midrule
macroeconomics & 5353 & 203.7 (540.85) & 58.0 & 165.0 & 0.135 \\
international & 3951 & 251.8 (681.66) & 75.0 & 224.0 & 0.171 \\
financial & 3953 & 207.8 (690.30) & 59.0 & 171.0 & 0.112 \\
economic history & 1061 & 125.9 (234.46) & 57.0 & 126.0 & 0.203 \\
microeconomic theory & 8079 & 109.4 (463.79) & 37.0 & 93.0 & 0.089 \\
industrial organization & 3985 & 167.5 (378.25) & 60.0 & 162.0 & 0.166 \\
econometrics \& quantitative & 6331 & 237.6 (872.89) & 66.0 & 175.0 & 0.128 \\
public & 3828 & 178.4 (398.10) & 68.0 & 169.0 & 0.196 \\
health & 1824 & 164.4 (325.80) & 78.0 & 175.0 & 0.230 \\
development & 2405 & 320.7 (715.11) & 132.0 & 317.0 & 0.308 \\
labor & 4946 & 230.7 (489.91) & 90.0 & 235.75 & 0.212 \\
agricultural & 928 & 151.1 (226.69) & 74.5 & 181.0 & 0.347 \\
urban \& regional & 933 & 246.8 (400.85) & 119.0 & 259.0 & 0.193 \\
environmental \& resource & 1520 & 175.9 (337.95) & 75.0 & 174.25 & 0.251 \\
behavioral & 2543 & 298.9 (786.60) & 95.0 & 257.5 & 0.091 \\
other & 1126 & 198.7 (579.87) & 58.0 & 161.75 & 0.256 \\
all & 52766 & 201.1 (587.07) & 65.0 & 176.0 & 0.163 \\
\bottomrule
\end{tabular}
    \caption{Descriptive Statistics by Economic Subfield. 
    ``Disruptive rate" is the fraction of papers with a disruption index greater than zero, following \citet{wu2019large}.}
    \label{tab:data_description}
\end{table}

\subsection{Semantic Content Extraction and Representation}
\label{sec:pipeline_for_embedding}
Our primary data consist of full-text research articles in PDF format. We developed a multi-stage pipeline to convert these documents into structured semantic representations suitable for calculating our isolation metrics.

First, we convert the PDFs to plain text using an LLM-based Optical Character Recognition (OCR) model. Second, we use LLMs to generate comprehensive summaries of each paper. A key innovation of our approach is the extraction of multiple, structured representations of each paper's intellectual content. Rather than relying on a single summary, we prompt the LLM to decompose each paper into five distinct semantic dimensions:
\begin{itemize}
\item \textbf{Paragraph:} A concise, high-level overview of the paper's content.
\item \textbf{Topic:} The paper's subfield, primary subject matter, and specific issues addressed.
\item \textbf{Question-Conclusion:} The core research questions and key findings.
\item \textbf{Methodology:} Data sources, analytical models, and technical approaches employed.
\item \textbf{Theory:} The theoretical frameworks, conceptual models, and analytical innovations.
\end{itemize}
This decomposition allows us to measure novelty along different dimensions of a paper's contribution - for example, distinguishing methodological novelty from theoretical novelty.

Third, to ensure the robustness of our semantic representations, we embed each of the five summary types using three complementary embedding models: SciBERT \citep{Beltagy2019SciBERT}, a BERT model pre-trained on a large corpus of scientific literature; SPECTER \citep{cohan2020specter}, a model specifically designed to generate document-level embeddings for scientific papers; and Qwen \citep{li2023towards}, a general-purpose state-of-the-art LLM. This triangulation approach mitigates the risk of model-specific biases.

Finally, we calculate pairwise distances between all paper embeddings using both Euclidean distance and cosine similarity. This yields a comprehensive feature space derived from the combination of five summary types, three embedding models, and two distance metrics. This systematic process results in 3240 candidate isolation features for each paper, providing a rich basis for our analysis. Further details are provided in Appendix \ref{app-sec:data}.

\paragraph{Addressing Look-Ahead Bias.}
The use of modern LLMs trained on vast internet corpora introduces a potential for ``look-ahead bias," where the model might inadvertently incorporate information unavailable at the time a paper was written \citep{sarkar2024lookahead}. We mitigate this concern through several strategies. 
First, our LLM task is summarization, not prediction. Different from prediction task such as predicting the citation count of a paper will receive by 2024 (which is likely to be in a LLM's training set), in our task the model is prompted to condense the provided text, not to infer future outcomes or context. 
Second, our prompts are designed to be content-focused, explicitly instructing the model to avoid external contextualization. 
Third, our metrics are based on relative semantic distances within a historical cohort, making them less sensitive to absolute shifts in the embedding space. 

\section{Validating Semantic Isolation Metrics}
\label{sec:prediction}

Before employing our semantic isolation metrics to construct a theoretical framework of novelty, we must first establish their empirical relevance. This section demonstrates that these metrics are powerful predictors of a paper's future impact. We show that they not only outperform established novelty measures but also provide complementary information, confirming that they capture a distinct and meaningful dimension of scientific contribution.

\subsection{Experimental Design}

We evaluate the predictive power of our isolation metrics through two supervised learning tasks: predicting long-term citation counts and identifying disruptive papers. Our experimental design is structured to test two key hypotheses: (1) that our isolation metrics have significant predictive power in isolation and (2) that their predictive content is not redundant with other textual or bibliometric features.
This validation strategy rests on a key premise: if our semantic isolation metrics capture meaningful dimensions of novelty, they should predict future impact beyond what is explained by existing measures and standard bibliometric controls. Strong predictive performance, particularly when competing against established novelty proxies like atypicality, provides evidence that our metrics identify papers positioned for significant influence.

We define several feature groups for our predictive models:
\begin{itemize}
\item \textbf{Controls:} Basic metadata including publication year, reference count, document length, word count, and sentence count.
\item \textbf{Subfield (Baseline):} Indicator variables for the LLM-classified subfields.
\item \textbf{Journal:} Journal-specific features including encoded journal identifiers and top-5 journal indicators.
\item \textbf{Textual Features:} Standard NLP features including word counts, part-of-speech tag frequencies, and other text-derived metrics.
\item \textbf{Topic Models:} Latent Dirichlet Allocation (LDA) and Term Frequency–Inverse Document Frequency (TF-IDF) representations. They contains future information that is not available at the time a paper is published, as they are calculated with the whole dataset.
\item \textbf{Atypicality:} The combination-based novelty scores developed by \cite{uzzi2013atypical}, including 10th percentile, median, and number of combinations, as provided by SciSciNet \citep{lin2023sciscinet}.
\item \textbf{Isolation:} Our proposed set of semantic isolation metrics, including point-in-time isolation metrics and retrospective dynamic isolation metrics\footnote{We do not include perspective dynamic isolation metrics in our main analysis. The version with perspective dynamic isolation metrics is in Appendix \ref{app-sec:temporal_isolation}.}.
\end{itemize}

To fully capture the nonlinear relationship between features, we employ XGBoost for prediction tasks. We use consistent hyperparameters across all feature combinations to ensure performance differences reflect information content rather than optimization artifacts. 
These hyperparameters are identified by tuning the model on the most complex feature set (the union of all feature groups), as parameters that perform well on this high-dimensional space are the most likely to be robust and generalize effectively to simpler feature subsets.
All results are averaged over 100 bootstrap iterations for robust statistical inference.

\subsection{Predicting Log Citation Counts}

We first assess the ability of our isolation metrics to predict long-term citation counts, a standard proxy for scientific impact. 
Our primary evaluation metric is the out-of-sample Mean Squared Error (MSE). We also report the "completeness score" from \citet{fudenberg2022measuring}, which measures the proportional reduction in MSE relative to a baseline feature set (Controls + Subfield). 
The larger the completeness (maximum is 1), the more error the tested model reduce from the base, which means the better the model's performance.
For example, using all features together gives a completeness score of 34.92\%, meaning that this large set of features can reduce 34.92\% of  our baseline prediction error. This suggests that predicting log citation counts is a hard task.



\paragraph{Results.}

Table \ref{tab:log_cite_individual} presents the standalone predictive power of the atypicality and isolation feature groups. 
When added to the control variables, the atypicality score offers no improvement over the baseline. In stark contrast, our isolation metrics achieve a completeness score of 9.93\%, demonstrating substantial predictive power. This is also surprising as only using isolation features can reach nearly 30\% of the completeness score of using all features (34.92\%), reassuring that isolation features predict citation count well.

\begin{table}[ht]
    \centering
\begin{tabular}{lll}
\toprule
Feature set & MSE & Completeness \\
\midrule
control+subfield (baseline) & 1.7044 & 0.0 \\
 & (0.00154) &  \\
control+atypicality & 1.7110 & -0.0039 \\
 & (0.00156) &  \\
control+isolation & 1.5352 & \textbf{0.0993}  \\
 & (0.00133) &  \\
\bottomrule
\end{tabular}
    \caption{Predicting log citation counts: individual feature groups. Standard errors from 100 bootstrap iterations in parentheses. Completeness is the proportional reduction in MSE relative to the baseline.}
    \label{tab:log_cite_individual}
\end{table}

\begin{table}[ht]
    \centering
\begin{tabular}{lll}
\toprule
Feature set & MSE & Completeness \\
\midrule
control+subfield & 1.7044 & 0.0 \\
 & (0.00154) &  \\
control+subfield+atypicality & 1.6646 & 0.0234 \\
 & (0.00154) &  \\
control+subfield+atypicality+isolation & 1.5090 & \textbf{0.1146} \\
 & (0.00133) &  \\
\midrule
all features (excl. isolation) & 1.1465 & 0.3273 \\
 & (0.00111) &  \\
all features (incl. isolation) & 1.1092 & \textbf{0.3492} \\
 & (0.00105) &  \\
\bottomrule
\end{tabular}
    \caption{Predicting log citation counts: additional value of isolation metrics.}
    \label{tab:log_cite_add_to_atypicality}
\end{table}

This result raises a key question: do isolation and atypicality measure the same underlying construct, with isolation being a more precise proxy? Or do they capture fundamentally different dimensions of novelty? Table \ref{tab:log_cite_add_to_atypicality} provides a clear answer. Adding atypicality to the baseline provides a modest 2.34\% improvement in completeness. However, subsequently adding our isolation metrics increases completeness to 11.46\% - an incremental gain of 9.12 percentage points. This strong additional effect provides compelling evidence that isolation and atypicality are complementary, measuring distinct aspects of novelty. Atypicality captures the novelty of a paper's inputs (unusual combinations of references), while our metrics capture the novelty of its output (the semantic content of its contribution).

Finally, we test whether the predictive power of our isolation metrics is robust to the inclusion of a comprehensive set of controls, including sophisticated textual representations like TF-IDF and LDA. The bottom panel of Table \ref{tab:log_cite_add_to_atypicality} shows that even when added to a model containing all other feature groups, our isolation metrics provide a statistically, significant improvement, increasing completeness by an additional 2.2 percentage points. This improvement is especially meaningful given that LDA and TF-IDF (in all features) contain future information that is not available at the time of the paper is published.

\begin{figure}[ht]
    \centering
    \includegraphics[width=0.5\linewidth]{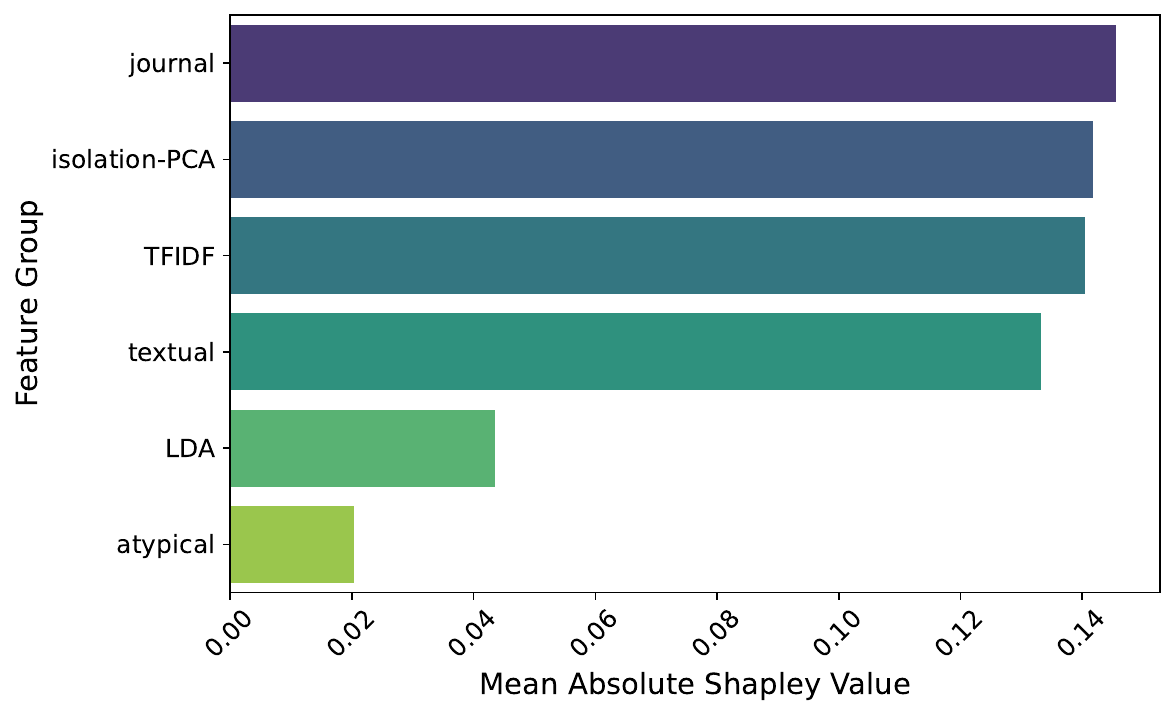}
    \caption{Feature importance (mean absolute SHAP) for predicting log citations.}
    \label{fig:log_cite_shap_cv}
\end{figure}

To assess the relative importance of each feature group, we compute SHAP (SHapley Additive exPlanations) values\footnote{We apply 5-fold cross validation, and combine the prediction on test folds to calculate the SHAP contribution.}. 
SHAP values provide a unified measure of feature importance by attributing each prediction to individual features. They are grounded in Shapley values from cooperative game theory, ensuring a fair and consistent allocation of contribution across features.
As shown in Figure \ref{fig:log_cite_shap_cv}, the isolation metrics (reduced to 200 dimensions via PCA for comparability) contribute more to the model's predictions than any other feature group, including TFIDF features. This confirms that semantic isolation is not a peripheral signal but a central predictor of scientific impact.

\subsection{Predicting Disruptive Papers}

We next turn to a more nuanced prediction task: identifying disruptive papers. This allows us to test whether our isolation metrics are merely a proxy for future attention (citations) or if they capture a more fundamental aspect of a paper's contribution. We frame this as a binary classification task, classifying a paper as disruptive if its Disruption Index (DI) is greater than zero \citep{wu2019large}.

Given the class imbalance (16.3\% of papers are disruptive), we focus on precision as our primary evaluation metric. Our goal is showing that semantic isolation is a high-quality signal that reliably predicting disruptive papers.
Precision answers the question: ``When our model predicts a paper is disruptive, how often is it correct?'' A high precision score provides strong evidence that semantic isolation is a reliable signal of disruptive potential. 
On the other hand, recall, which measures  ability to identify every disruptive paper, is of secondary importance. 
We also report the F1-score and the completeness score (where error is defined as ($1-precision$).
We use balanced class weights during training and report mean F1 and precision across 100 times bootstrapping.

\begin{table}[ht]
    \centering
\begin{tabular}{lllll}
\toprule
Feature set & Precision & Comp-prec & F1 & Comp-F1 \\
\midrule
control+subfield & 0.3140 & 0.0 & 0.4151 & 0.0 \\
 & (0.00053) &  & (0.0057) &  \\
control+atypicality & 0.3231 & 0.0133 & 0.4220 & 0.0118 \\
 & (0.00057) &  & (0.00058) &  \\
control+isolation & 0.4064 & \textbf{0.1347} & 0.4699 & 0.0937 \\
 & (0.00064) &  & (0.00059) &  \\
\bottomrule
\end{tabular}
    \caption{Predicting disruptive papers: individual feature groups. Standard errors from 100 bootstrap iterations in parentheses. Comp-prec shows the completeness of error in precision, and Comp-F1 shos the completeness of error in F1.}
    \label{tab:DI_paper_individual}
\end{table}

\paragraph{Results.}
The results for disruption prediction, presented in Table \ref{tab:DI_paper_individual}, mirror the patterns observed for citations. Our isolation metrics substantially outperform the atypicality score, achieving a 13.47\% completeness score on precision compared to just 1.33\% for atypicality. This consistency across different prediction tasks strengthens the evidence that isolation captures fundamental aspects of research novelty.

Moreover, the metrics are again complementary, as shown in Table~\ref{tab:DI_paper_add_to_atypicality}. Adding isolation metrics to a model that already includes the baseline and atypicality features increases precision completeness from 4.34\% to 15.13\%. This additional effect confirms that our measure of semantic distance captures a dimension of novelty relevant for disruption that is distinct from the recombinative novelty captured by atypicality.

\begin{table}[ht]
    \centering
\begin{tabular}{lllll}
\toprule
Feature set & Precision & Comp-prec & F1 & Comp-F1 \\
\midrule
control+subfield & 0.3140 & 0.0 & 0.4151 & 0.0 \\
 & (0.00053) &  & (0.00057) &  \\
control+subfield+atypicality & 0.3437 & 0.0434 & 0.4424 & 0.0466 \\
 & (0.00054) &  & (0.00051) &  \\
control+subfield+atypicality+isolation & 0.4178 & \textbf{0.1513} & 0.4788 & 0.1089 \\
 & (0.00072) &  & (0.00064) &  \\
\midrule
all features (excl. isolation) & 0.4270 & 0.1647 & 0.4915 & 0.1305 \\
 & (0.00070) &  & (0.00060) &  \\
all features (incl. isolation) & 0.4517 & \textbf{0.2007} & 0.5063 & 0.1559 \\
 & (0.00081) &  & (0.00067) &  \\
\bottomrule
\end{tabular}
    \caption{Predicting disruptive papers: additional value of isolation metrics. Comp-prec shows the completeness of error in precision, and Comp-F1 shos the completeness of error in F1.}
    \label{tab:DI_paper_add_to_atypicality}
\end{table}

The bottom panel of Table \ref{tab:DI_paper_add_to_atypicality} shows that isolation metrics continue to provide significant predictive value even in the presence of all other features, increasing precision completeness from 16.47\% to 20.07\%. 
This 3.6 percentage point improvement demonstrates the consistent and non-redundant value of isolation across different impact dimensions.

\begin{figure}[ht]
    \centering
    \includegraphics[width=0.5\linewidth]{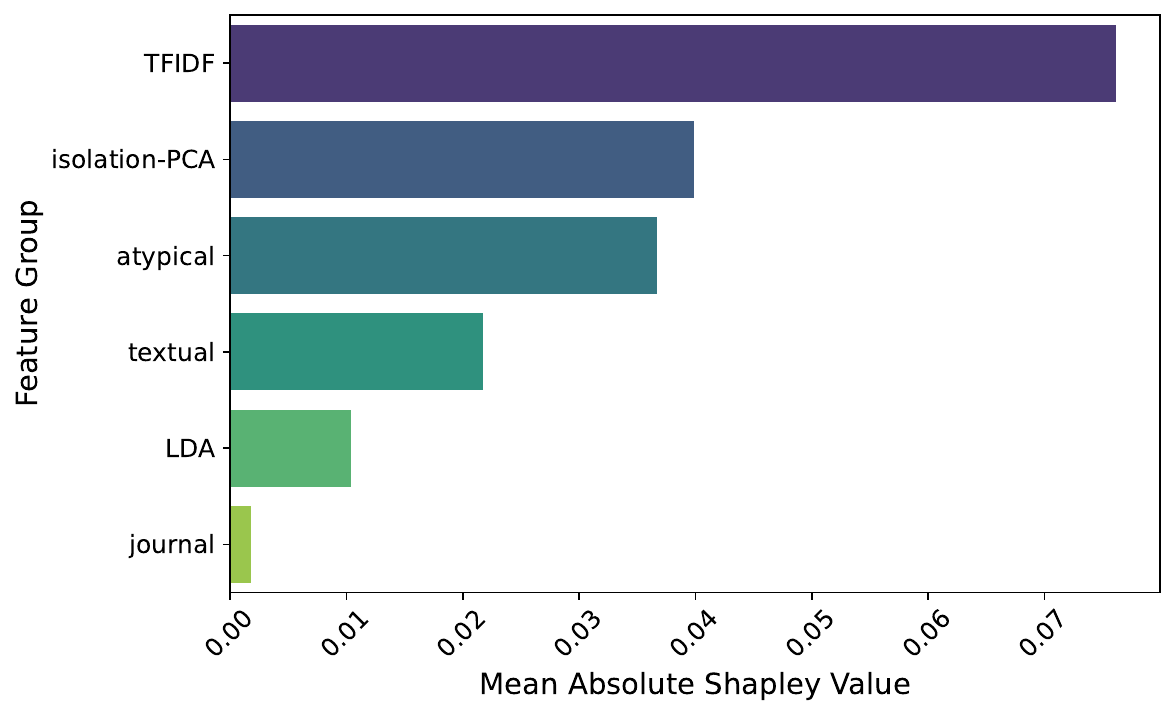}
    \caption{Feature importance (mean absolute SHAP) for predicting disruption.}
    \label{fig:DI_paper_shap_cv}
\end{figure}

The SHAP analysis\footnote{We apply 5-fold cross validation, and combine the prediction on test folds to calculate the shap contribution.} in Figure \ref{fig:DI_paper_shap_cv} further confirms that our isolation features are a key driver of the model's predictions.
This confirms that isolation captures meaningful signals for identifying potentially disruptive research, further validating our approach across different dimensions of research impact.

\subsection{Summary of Validation}
Our predictive analyses provide robust evidence for the empirical relevance between our semantic isolation metrics and future impacts. Across two distinct impact measures - citations and disruption - these metrics demonstrate substantial predictive power, significantly outperforming the established atypicality benchmark. Crucially, their predictive content is both complementary to atypicality and non-redundant with a comprehensive set of textual and bibliometric controls, confirming that they capture a novel dimension of scientific contribution rooted in a paper's semantic positioning and temporal dynamics. Notably, these features are calculated with information available at the publication time of a paper, making it a useful tool to assess the future impact at a paper's early stage.

As detailed in Appendix \ref{app-sec:isolation_specific_group}, these findings are robust across economic subfields. 
Decomposition (see Appendix \ref{app-sec:isolation_insight}) reveals that features related to papers' core contributions (paragraph summaries and question-conclusion framing) provide the strongest predictive signals.
Appendix \ref{app-sec:temporal_isolation} shows that including prospective metrics can significantly improve the predicting performance, serving as a validation of our temporal novelty concept.

Having established the empirical relevance between our semantic isolation metrics and impacts, we now proceed to use them as the measurement foundation for our two-dimensional framework of novelty.

\section{A Two-Dimensional Framework of Novelty}
\label{sec:typology_novelty}

Having established that our semantic isolation metrics robustly predict scientific impact, we now employ them as the measurement foundation for our two-dimensional framework of novelty.
This section moves beyond treating novelty as a monolithic concept, demonstrating how distinct pathways to intellectual contribution can be empirically measured and systematically categorized.

\subsection{Constructing Spatial and Temporal Novelty Scores}
\label{sec:construct_novelty_score}
We operationalize our framework by constructing two composite scores that correspond to the conceptual dimensions of novelty outlined previously. 
These scores are calculated using only information available at the time of publication, ensuring clear interpretation of temporal relationships.

While we have a large set of semantic isolation features, we need to derive scalar values for spatial novelty and temporal novelty. Simple aggregation methods such as taking the mean are straightforward but would lose the nuances captured by individual features. For example, averaging the 30-NN distance and 3-NN distance would either be meaningless or dominated by the larger 30-NN values. Therefore, we employ Principal Component Analysis (PCA) to preserve most of the variation in the features while reducing their dimensionality.

\begin{itemize}
\item \textbf{Spatial Novelty Score:} We apply PCA to the set of contemporaneous point-in-time isolation metrics (i.e., metrics calculated using all prior work, $t=0$) and the baseline paper-level mean distance\footnote{Basic features of semantic distance, introduced in Appendix \ref{app-sec:baseline_semantic_metric}.}. The first principal component serves as our spatial novelty score. This score captures a paper's static semantic isolation from the existing body of knowledge at its moment of publication.
\item \textbf{Temporal Novelty Score:} We apply PCA to the set of retrospective dynamic isolation metrics, which measure the change in a paper's semantic neighborhood in the years leading up to its publication (e.g., the change from $t=-3$ to $t=0$). The first principal component serves as our temporal novelty score, capturing the degree to which a paper is situated within a rapidly evolving or "hot" research frontier.
\end{itemize}

The use of contemporaneous point-in-time metrics and  retrospective dynamic metrics makes sure that our framework are not using any future information. By using the first principal component for each, we create summary indices that capture the largest shared dimension of variation across our granular metrics, reducing measurement noise while retaining the core signal of each novelty type. Further details on the PCA construction are provided in Appendix \ref{app-sec:PCA}.

\subsection{A Typology of Research Neighborhoods}
We classify each paper based on the state of its semantic neighborhood, defined by its position in the two-dimensional novelty space. A neighborhood is classified as "High" on a given dimension if the paper's score on that dimension is above the median. Crossing these two classifications yields a four-quadrant typology, with each quadrant representing a distinct archetype of research environment:

\begin{itemize}
    \item \textbf{Consolidating Neighborhoods (Low Spatial / Low Temporal):} These are stable, well-established research areas where contributions are semantically close to existing work. Papers in these neighborhoods typically refine or incrementally extend established paradigms.
    \item \textbf{Outlying Neighborhoods (High Spatial / Low Temporal):} These are intellectually distant and slow-moving or static research areas. Papers here represent unconventional ideas disconnected from contemporary research trends.
    \item \textbf{Trendy Neighborhoods (Low Spatial / High Temporal)}: These are rapidly evolving research frontiers where contributions, while not semantically distant from their immediate peers, are part of a concentrated wave of scientific interest.
    \item \textbf{Trailblazing Neighborhoods (High Spatial / High Temporal)}: These are nascent, intellectually distinctive areas of inquiry that are also rapidly gaining momentum. Papers in these neighborhoods are positioned at the start of new, fast-growing research directions.
\end{itemize}

It is crucial to emphasize that this typology characterizes the context in which a paper is published, not the intrinsic attributes of the paper itself. A paper is thus described by the type of neighborhood it inhabits (e.g., "a paper in a Trailblazing neighborhood"). This distinction is fundamental to our analysis, as it shifts the focus from a paper's identity to its strategic position within the scientific landscape.

\begin{figure}[h!]
    \centering
    \includegraphics[width=0.8\linewidth]{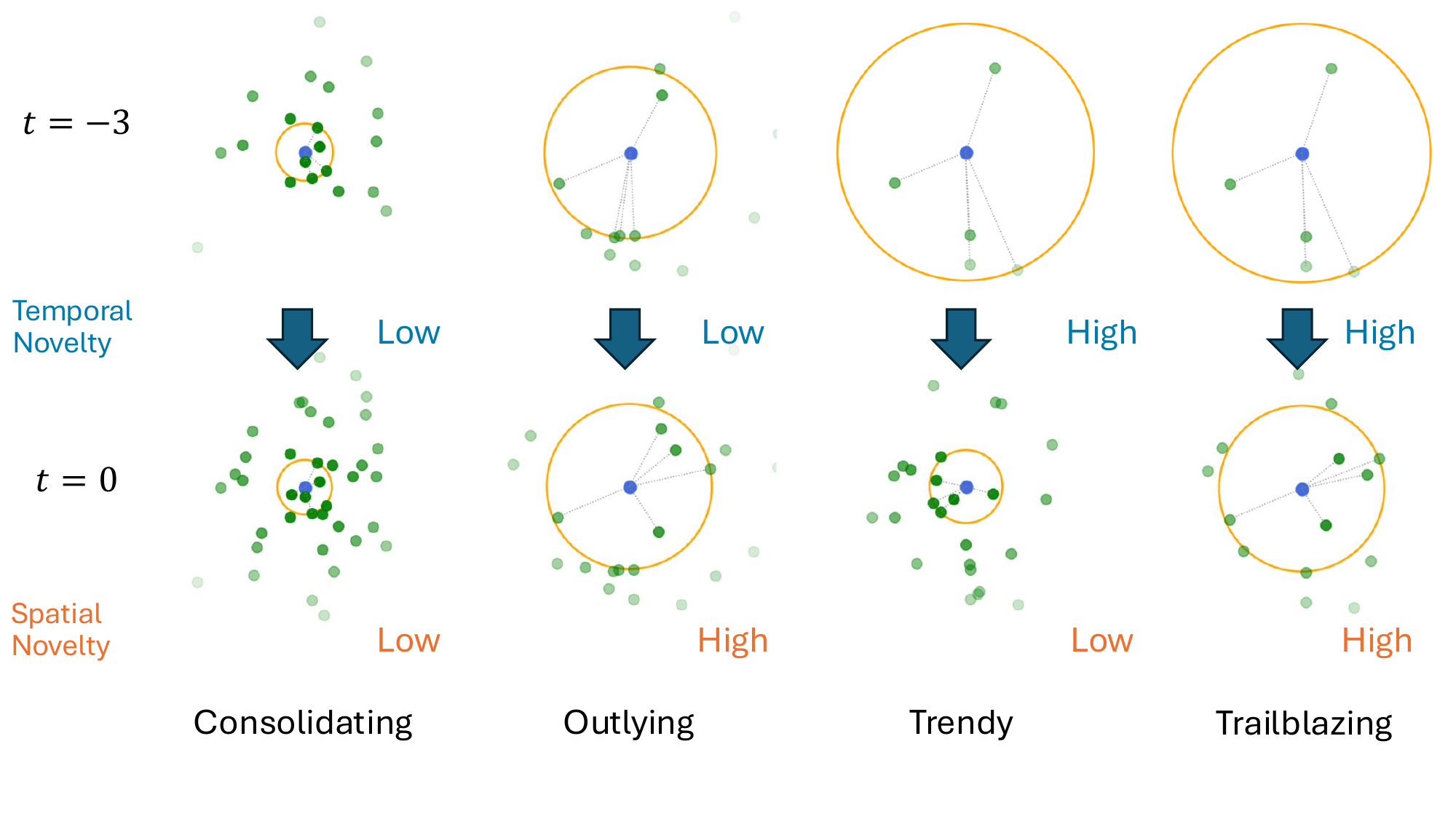}
    \caption{
    A stylized illustration of the four novelty archetypes. 
    The blue point is the focal paper; green points are prior work. The orange circle represents the paper's isolation level (e.g., its 5-NN distance) at publication ($t=0$) and three years prior ($t=-3$).
    High spatial novelty is indicated by a large radius at $t=0$.
    High temporal novelty is indicated by a large change in the radius between $t=-3$ and $t=0$.
    }
    \label{fig:illustration_four_types}
\end{figure}

Figure \ref{fig:illustration_four_types} provides a stylized illustration of these four archetypes in semantic space. High spatial novelty corresponds to a large isolation radius at the time of publication ($t=0$). 
High temporal novelty is visualized by a significant decrease in the isolation radius between a prior period ($t=-3$) and the publication date, indicating a rapid densification of the paper's intellectual neighborhood.

A key feature of our framework is the distinction between static isolation and its rate of change. One might consider using spatial novelty at $t=0$ and spatial novelty at $t=-3$ as the two dimensions. However, for a paper with high spatial novelty, its novelty three years prior is also likely to be high, providing little additional information. Our temporal novelty score, by measuring the change in isolation, is designed to capture this dynamic dimension directly. It effectively differentiates between an isolated paper in a static field (Outlying neighborhoods) and an isolated paper in a field that is rapidly forming around it (Trailblazing neighborhoods).

\subsection{The Structure of the Novelty Space}
An analysis of the joint distribution of the two novelty scores reveals important structural features of the scientific landscape.

\begin{figure}[h!]
    \centering
    \includegraphics[width=0.5\linewidth]{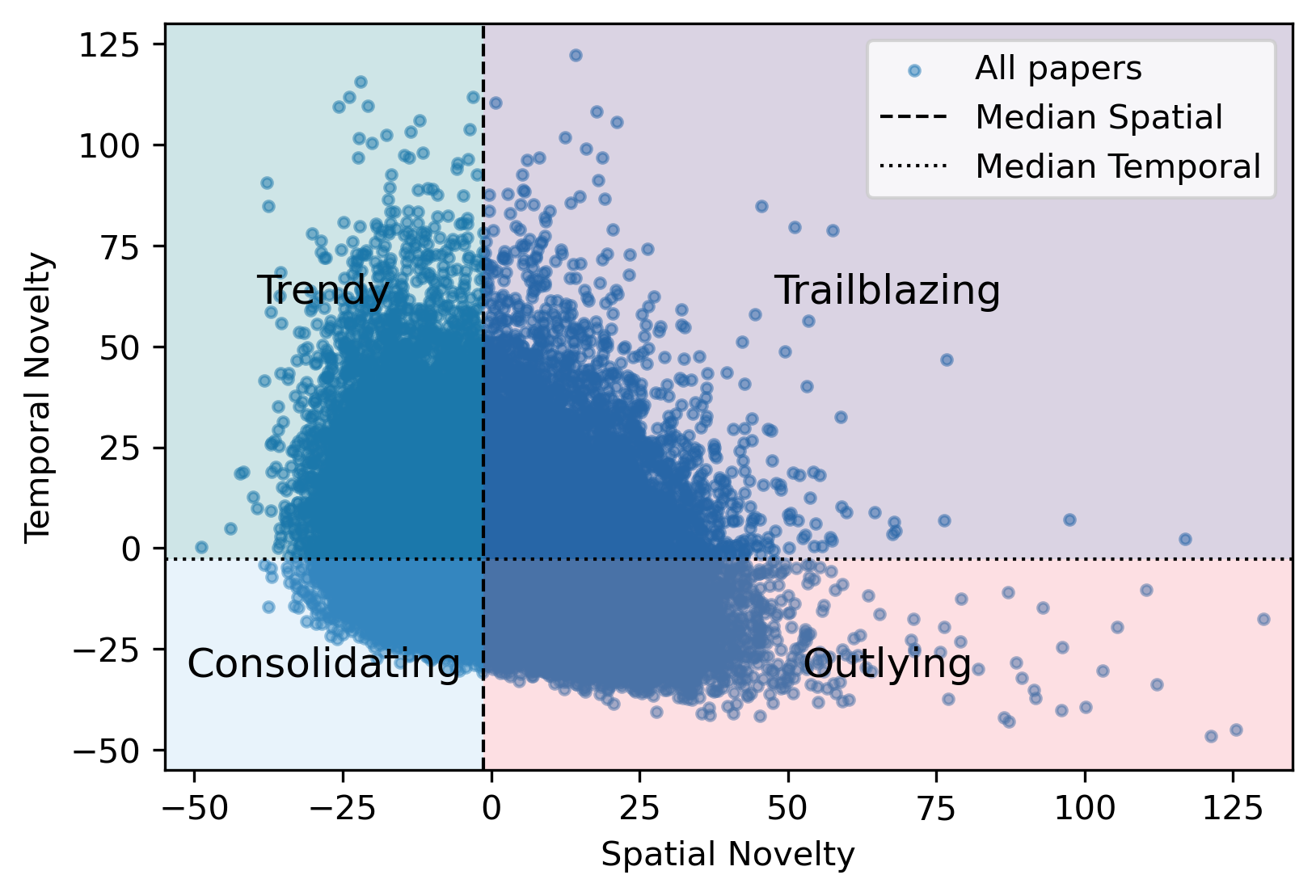}
    \caption{The joint distribution of spatial and temporal novelty. The negative correlation suggests a structural trade-off.}
    \label{fig:quadrant_paper_scatter}
\end{figure}

\paragraph{A Structural Trade-Off.}
The spatial and temporal novelty scores exhibit a statistically significant negative correlation $r=-0.26$. This finding provides quantitative evidence of a structural trade-off between the two pathways to novelty. 
In other words, it is empirically challenging for a single work to be both radically different from prior art (high spatial novelty) and situated at the heart of a burgeoning research trend (high temporal novelty).
This trade-off is reflected in the distribution of papers across our typology (Figure \ref{fig:quadrant_paper_scatter}). The off-diagonal quadrants --- Trendy and Outlying neighborhoods --- are the most populated (each containing about 30\% of papers), representing the more common research strategies. The relative scarcity of papers in the Trailblazing neighborhoods (about 20\%) underscores the difficulty of simultaneously achieving high spatial and temporal novelty.
This negative correlation points to a stylized fact of scientific production: a structural trade-off exists between intellectual distinctiveness and timely engagement

\paragraph{Mapping Economic Subfields.}

\begin{figure}[h!]
    \centering
    \includegraphics[width=0.8\linewidth]{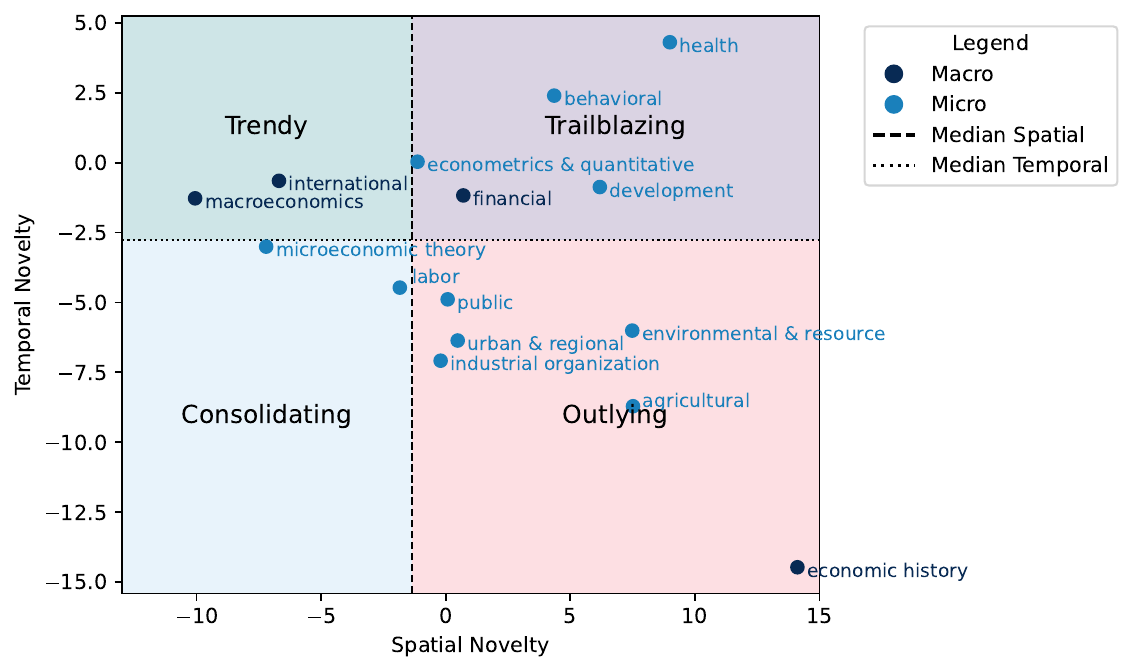}
    \caption{Median novelty profiles by economic subfield. Each point represents the median spatial and temporal novelty score for all papers in a given subfield. 
    }
    \label{fig:field_novelty_sem_median}
\end{figure}

Figure \ref{fig:field_novelty_sem_median} plots the median novelty scores for each economic subfield. The patterns align with what one might expect given these fields' methodological approaches. Economic History, with its emphasis on context-specific analysis, falls predominantly into the Outlying quadrant. Microeconomic Theory, building on a stable set of foundational models, centers in the Consolidating quadrant. Fields associated with recent rapid growth, such as Behavioral and Health Economics, appear in the Trailblazing quadrant.

While different fields tend to cluster in different quadrants, these mappings reflect only median tendencies and mask considerable within-field heterogeneity. Individual papers within every field span multiple archetypes. For instance, Behavioral Economics --- though exhibiting higher novelty on average --- includes substantial Consolidating work that refines foundational insights alongside Trailblazing explorations of new domains. This diversity underscores that our framework characterizes research strategies, not research quality: each quadrant represents a legitimate scholarly approach with distinct contributions to knowledge accumulation.
Importantly, our framework also reveals substantial overlap across fields with distinct median profiles. 
Appendix \ref{app-sec:descriptive_stat} presents both the inter-quartile range version and the full distributions for Macroeconomics and Economic History, confirming that our novelty scores identify research archetypes that transcend field boundaries, and provides detailed distributions for all subfields that substantiate the structural trade-off and field-level patterns.


\section{Novelty and Impacts}
\label{sec:novelty_impact}
Having established our two-dimensional novelty framework, we now test its central prediction: that different pathways to novelty produce systematically different forms of scientific impact. We examine whether our four neighborhood archetypes predict differential patterns in two impact outcomes --- high citation rates and disruptive influence. To ensure robustness, all results are averaged over 100 bootstrap iterations\footnote{We provide standard errors in Appendix \ref{app-sec:novelty_impact}.}.

\subsection{Temporal Novelty Predicts High Citations}
\label{sec:novelty_impact_high_cite}
\begin{figure}[h!]
    \centering
    \includegraphics[width=0.5\linewidth]{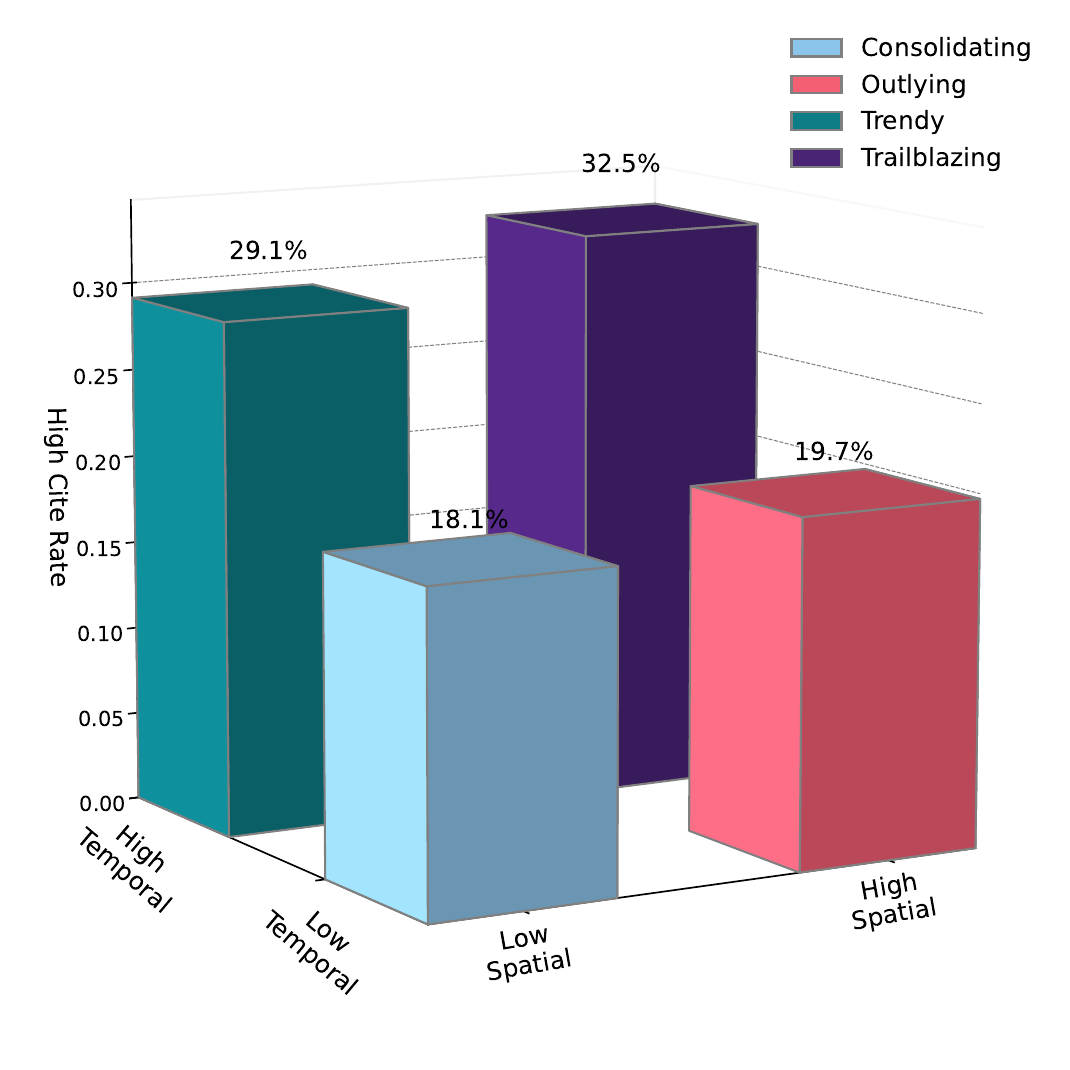}
    \caption{Probability of Being a Highly Cited Paper (Top 25\%) by Novelty Type. The baseline probability of being a top-25\% cited paper is 24.77\%. }
    \label{fig:quadrant_top25_cite_rate}
\end{figure}

Our first analysis examines the likelihood of a paper becoming highly cited, defined as reaching the top 25\% of citations within its publication-year-subfield cohort. The results, presented in Figure \ref{fig:quadrant_top25_cite_rate}, reveal a clear and powerful pattern: temporal novelty is the primary predictor of citation success.

Increasing temporal novelty from low to high dramatically increases a paper's probability of being highly cited. Among papers with low spatial novelty, papers in Trendy neighborhoods are 1.6 times more likely to be highly cited than papers in Consolidating neighborhoods (29.1\% vs. 18.1\%). Similarly, among papers with high spatial novelty, papers in Trailblazing neighborhoods are 1.65 times more likely than papers in Outlying neighborhoods (32.5\% vs. 19.7\%). This high-temporal-novelty advantage of over 10 percentage points is substantial and consistent across both levels of spatial novelty.

In contrast, the effect of spatial novelty is modest. Moving from low to high spatial novelty increases the high-citation probability by only 1.6 percentage points for low-temporal-novelty papers and 3.4 percentage points for high-temporal-novelty papers. While statistically significant\footnote{Detail numbers appear in Table \ref{app-tab:quadrant_top25_cite_rate_table}, Appendix \ref{app-sec:high_cite_rate_quadrant}.}, this effect is an order of magnitude smaller than that of temporal novelty.

This pattern suggests that positioning a contribution within a dynamic research frontier confers a powerful attentional advantage. Papers that engage with rapidly evolving conversations benefit from the field's momentum, the influx of new researchers, and the dense citation networks that form around emerging topics.

\subsection{Spatial Novelty Predicts Disruption}
\label{sec:novelty_impact_disruption}
\begin{figure}[h!]
    \centering
    \includegraphics[width=0.5\linewidth]{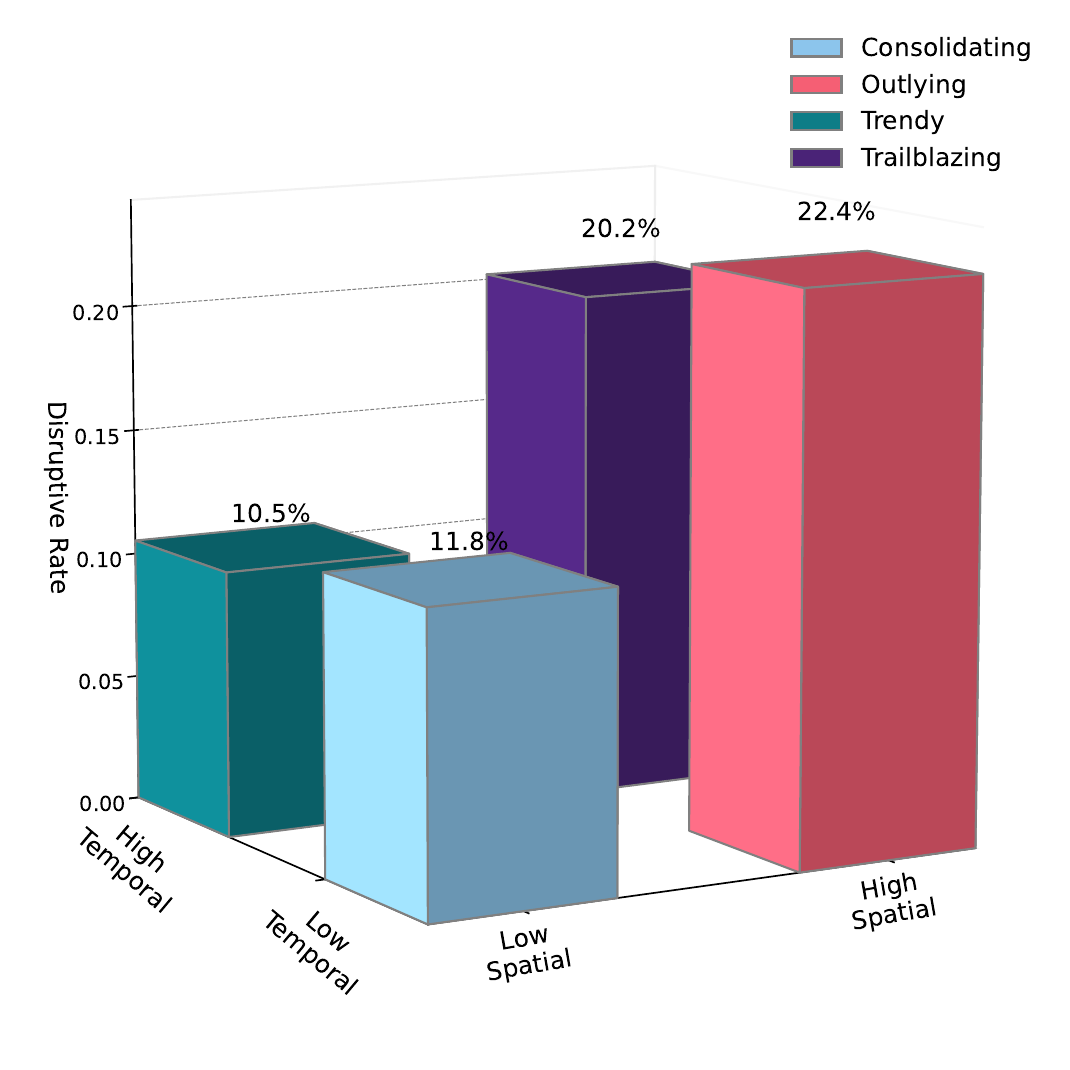}
    \caption{Probability of being a disruptive paper by novelty type. The baseline probability of being disruptive is 16.30\%.}
    \label{fig:quadrant_disruptive_rate_all}
\end{figure}

The relationship between novelty and impact reverses when we examine disruption. As shown in Figure \ref{fig:quadrant_disruptive_rate_all}, spatial novelty emerges as the dominant predictor of disruptive influence.

Increasing spatial novelty from low to high nearly doubles a paper's probability of being disruptive. Papers in Outlying neighborhoods are 1.9 times more likely to be disruptive than papers in Consolidating neighborhoods (22.4\% vs. 11.8\%), and papers in Trailblazing neighborhoods are 1.9 times more likely than papers in Trendy neighborhoods (20.2\% vs. 10.5\%). This high-spatial-novelty advantage is large and consistent across both levels of temporal novelty.

Conversely, temporal novelty is weakly negatively associated with disruptive potential.
High-temporal-novelty papers are approximately 1.3 percentage points less likely to be disruptive than their low-temporal-novelty counterparts. This finding is conceptually intuitive: disruption, by definition, involves rendering prior work obsolete. Such a stark break is more likely to occur when a paper introduces a truly distinct perspective (high spatial novelty) rather than when it participates in a crowded, cumulative conversation at an evolving frontier (high temporal novelty).


\subsection{Navigating the Trade-off: The Trailblazing Archetype}
\label{sec:pioneer_advantage}
Our framework identifies Trailblazing as a unique archetype that successfully navigates the trade-off between spatial and temporal novelty. While these are the rarest neighborhoods, their impact profile is exceptional. Papers in Trailblazing neighborhoods are not only strong performers on both individual impact dimensions (32.5\% high-citation rate, 20.2\% disruption rate), but they also exhibit a disproportionate likelihood of achieving both forms of impact simultaneously.

\begin{figure}[h!]
    \centering
    \includegraphics[width=0.5\linewidth]{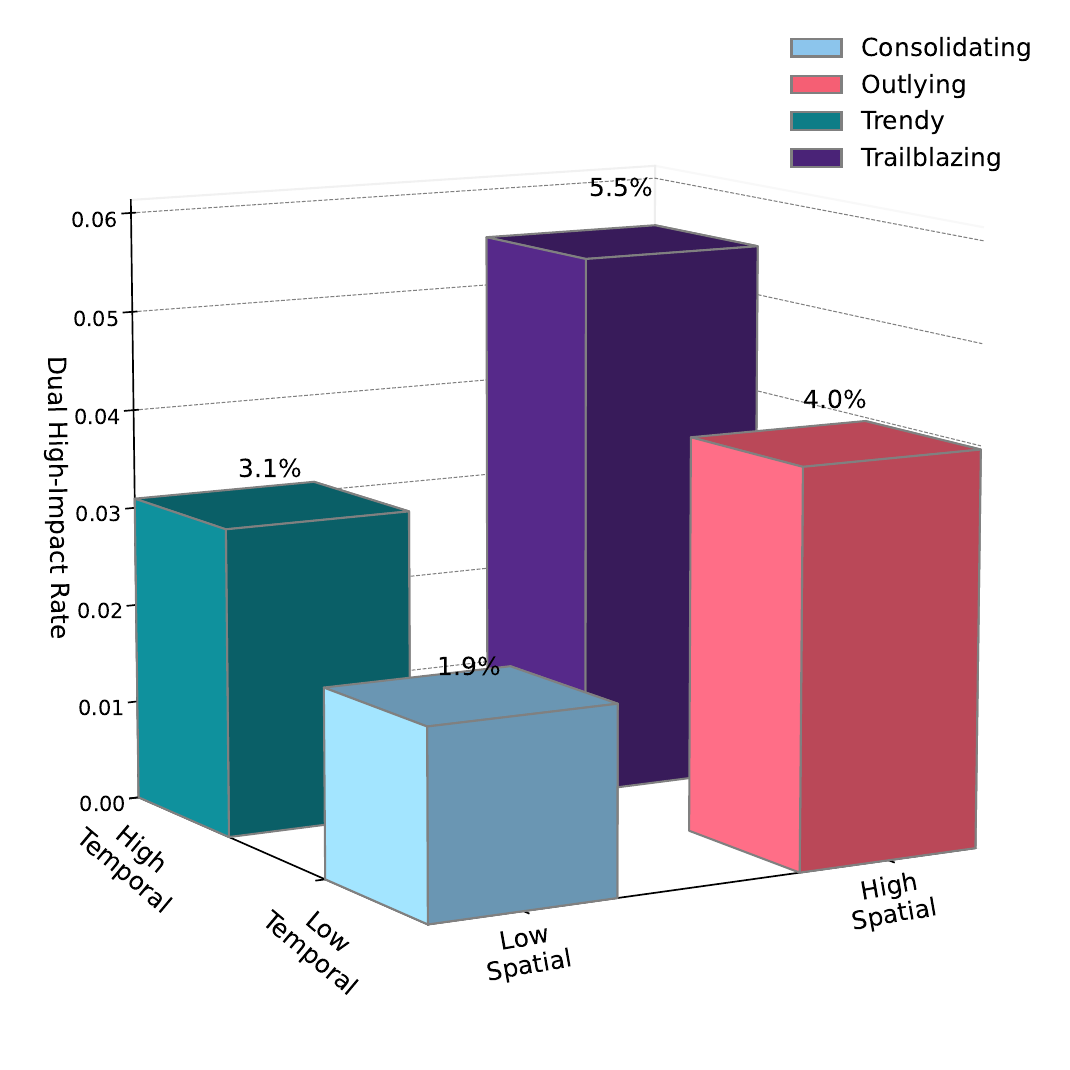}
    \caption{Fraction of papers that are both top 25\% high-cited and disruptive, categorized by novelty type. The base rate of being both high-cited and disruptive is 3.59\%.}
    \label{fig:quadrant_disruptive_and_top25_rate}
\end{figure}

As shown in Figure \ref{fig:quadrant_disruptive_and_top25_rate}, Papers in Trailblazing neighborhoods have a 5.5\% probability of being both highly cited and disruptive. This is nearly three times the rate of papers in Consolidating neighborhoods (1.9\%) and significantly higher than that of papers in Outlying are (4.0\%) or Trendy neighborhoods (3.1\%). 
This high dual-impact probability suggests a powerful synergistic effect: being both intellectually distinct (spatial) and well-timed (temporal) does not merely add to a paper's impact, but amplifies its transformative potential, creating a rare pathway to both immediate relevance and lasting influence.

\subsection{Robustness and Alternative Specifications}
The core patterns identified in our analysis prove robust across a wide range of alternative specifications. As detailed in the appendix, the dominance of temporal novelty for achieving high-citation status and of spatial novelty for generating disruptive influence holds when using stricter citation thresholds (top 10\%, top 1\%) and when examining disruption rates within different subsamples of highly-cited papers (Appendices \ref{app-sec:high_cite_rate_quadrant} and \ref{app-sec:quadrant_disruptive_rate}). 
An analysis of median citation counts adds further nuance: while confirming the primary role of temporal novelty, it also shows that spatial novelty provides a significant positive boost to citation levels, particularly for papers already high in temporal novelty (Appendix \ref{app-sec:novelty_and_citation_counts}).
The synergistic advantage in achieving dual impact is similarly robust for papers in Trailblazing neighborhoods (Appendix \ref{app-sec:quadrant_high_cite_and_disruptive_rate}).

Furthermore, the framework's integrity is confirmed by its insensitivity to key methodological choices. The fundamental relationships are preserved when using a more restrictive 75th percentile threshold to define our novelty archetypes (Appendix \ref{app-sec:split_high_75}). Crucially, field-by-field analyses demonstrate that the observed patterns hold within subfields, confirming that our findings represent general pathways to scientific influence rather than mere artifacts of disciplinary differences (Appendix \ref{app-sec:field_level}).

This body of evidence strongly supports our multidimensional framework. It challenges unidimensional approaches to novelty assessment by demonstrating that the pathways to scientific impact are not singular but plural, each requiring distinct intellectual strategies and offering different rewards.

\section{Discussion}
\label{sec:discussion}
This study challenges the monolithic view of scientific novelty by developing and validating a multidimensional framework that distinguishes between fundamentally different pathways to impact. Our analysis of the economics literature reveals that novelty is best understood not as an intrinsic property of a paper, but as a relational construct defined by a paper's position in the intellectual landscape. We show this position operates along two distinct and measurable dimensions --- spatial and temporal --- that systematically produce different forms of scientific influence.

\subsection{Core Findings and Implications}
Our analysis yields three primary discoveries that reshape the understanding of the novelty-impact relationship.

First, novelty is a multidimensional construct with a clear structural trade-off. We demonstrate that spatial novelty (intellectual distance from prior work) and temporal novelty (engagement with a dynamic research frontier) are empirically distinct and negatively correlated. This reveals a fundamental tension in research strategy: it is empirically difficult to be simultaneously highly original and perfectly timed within a burgeoning field.

Second, different dimensions of novelty predict different forms of impact. Temporal novelty is the primary predictor of citation success, as it positions a paper within an active conversation that attracts community attention. Spatial novelty, in contrast, is the primary predictor of disruptive influence, as it provides the intellectual distance necessary to render prior approaches obsolete. These are not interchangeable measures of ``being novel'' but distinct positioning patterns associated with different impact profiles.

Third, we identify a synergistic advantage associated with papers in Trailblazing neighborhoods. Papers that successfully navigate the trade-off to achieve both high spatial and high temporal novelty are disproportionately likely to achieve both high citations and high disruption. This suggests that distinctiveness and timeliness are not merely additive but interact to amplify a paper's transformative potential.

Underlying these empirical findings is a central theoretical contribution: the decoupling of a paper's identity from its position. Traditionally, novelty is treated as an intrinsic attribute of a contribution. Our framework reframes it as a positional good, defined by a paper's relationship to its surrounding semantic neighborhood. 
This conceptual shift from an internal property to an external context is crucial. It moves the unit of analysis from the paper itself to the research locale it inhabits, allowing us to study how different types of intellectual positioning relate to different forms of scientific influence --- how researchers select into and navigate different types of intellectual environments.
This perspective provides a new mechanism for understanding scientific progress, one that accounts for both individual contributions and the ecological dynamics of the knowledge landscape in which they are embedded.

Our framework offers new tools for science policy and evaluation. By distinguishing between citation-generating and disruption-generating forms of novelty, our metrics reveal that current citation-heavy evaluation systems may systematically underweight the contributions of Outlying work. Papers in Outlying neighborhoods exhibit high disruptive potential but lower citation rates, suggesting their transformative value may be less visible through conventional metrics. Our framework provides a complementary lens for identifying different novelty profiles, enabling more nuanced assessment of what constitutes a ``transformative'' contribution across different impact dimensions.


Ultimately, our framework calls for a fundamental shift in how scientific contributions are discussed and evaluated. The common question, ``Which papers are most novel?'' is ill-posed. Our findings suggest it should be replaced with: ``What type of novelty does this paper exhibit?'' This reframing acknowledges that excellence in science takes multiple forms - each valuable, each measurable, and each essential for the collective progress of a thriving scientific enterprise.

\subsection{Limitations and Future Research}
Our study is subject to several limitations that point toward important directions for future research.

\paragraph{Causal inference and unobserved heterogeneity.} A primary limitation of our study is that our findings are associational, not causal. It is plausible that unobserved factors, such as innate researcher talent (a 'genius effect') or external shocks like new funding opportunities, could drive both a paper's positioning in a particular neighborhood and its eventual impact. While our framework's ability to decouple a paper's position from its intrinsic qualities offers a theoretical pathway to mitigate this, future work is needed to more rigorously identify causal effects. For instance, quasi-experimental designs leveraging exogenous shocks to specific research fields, combined with the inclusion of author-level fixed effects, could help disentangle the effect of strategic positioning from that of unobserved talent.

\paragraph{Measurement, comparability, and generalizability.} Our framework relies on the comparability of semantic distances, a challenge given the heterogeneity within economics. Our primary approach treats economics as a single intellectual space, applying a global normalization to our novelty metrics. This choice is theoretically motivated by the shared language and intellectual heritage across economic subfields, allowing us to capture valuable cross-subfield connections. We validate this approach in the appendix, where field-by-field analyses demonstrate that our core findings --- the divergent impacts of spatial and temporal novelty --- are not driven by any single subfield but represent a general dynamic within economics.

However, extending this framework to compare fundamentally different disciplines (e.g., economics versus computer science) would require a more nuanced approach. A global comparison might be confounded by baseline differences in disciplinary density and structure. A promising direction for future cross-disciplinary research would be to employ a two-stage normalization: first, embedding all papers in a unified semantic space to preserve global relationships, but then normalizing novelty scores and defining archetypes relative to each discipline's own distribution. This hybrid approach would enable meaningful comparisons of innovation patterns across fields, a key priority for developing a general theory of scientific progress. This extension, along with incorporating pre-print archives and non-journal publication outlets, will be crucial for testing the full generalizability of our findings.

\paragraph{Mechanisms of impact.} While we identify that different novelty types lead to different impacts, our study does not fully unpack the why. The mechanisms driving these patterns --- be it reviewer bias, reader attention dynamics, or the structure of citation networks --- remain to be fully explored. Future work could leverage citation context analysis or even experimental designs to disentangle these channels, moving from a macro-level map of scientific impact to a micro-level understanding of its underlying drivers.

\section{Conclusion}
\label{sec:conclusion}

This paper challenges a monolithic view of novelty, proposing instead a multidimensional framework that distinguishes a paper's intellectual distinctiveness (spatial novelty) from its engagement with an emerging frontier (temporal novelty). We develop a scalable toolkit to measure these dimensions and provide robust evidence that they are tied to fundamentally different forms of scientific impact: citations and disruption.
Our central finding is that the pathway to scientific influence is not singular, but plural. Recognizing these diverse pathways has direct implications for science policy and evaluation. Funding and promotion systems that rely heavily on citation-based metrics risk systematically favoring contributions in well-timed, Trendy neighborhoods at the expense of intellectually distant, Outlying work with disruptive potential. Our framework offers a more nuanced vocabulary and a practical toolkit for cultivating a more diversified and ultimately more innovative research portfolio.
The conceptual shift proposed here --- from novelty as an intrinsic attribute to a positional good --- opens up numerous avenues for future research. It invites a new program of inquiry focused on the strategic dynamics of scientific discovery: 
how do these trade-offs manifest across researchers, institutions, and funding contexts?
By providing a quantitative map of the intellectual landscape, our work lays the groundwork for a richer, more ecological understanding of how science progresses.


\clearpage

\bibliographystyle{ecta}
\bibliography{reference}

\clearpage

\appendix

\section{Technical Details of Semantic Isolation Metrics}
\label{app-sec:isolation_indicators}

This appendix provides the formal definitions for the semantic isolation metrics discussed in Section \ref{sec:framework}.

\subsection{Fundamental Definitions}
Let $S$ be the corpus of all papers. For each paper $i\in S$, we have a vector embedding $e_i$ and a publishing year $t_i$. Let $S(t)=\{j\in S \mid t_j\leq t \}$ be the set of papers published in or before year $t$.

The core of our measurement is the set of pairwise distances between a focal paper $i$ and and a specified comparison set of papers. For a focal paper $i$ and a temporal offset $c$ (in years), we define the comparison set as $S(t_i+c)$. The set of distances between paper $i$ and all other papers $j\in S(t_i+c)$ (where $j\neq i)$ is given by:
\[
D_{\{i, c\}} =\{ d(e_i, e_j) \mid j\in S(t_i+c), ~j\neq i \}
\]
where $d(\cdot)$ denotes a measure of distance (e.g., Euclidean distance). The offset $c$ defines the temporal window of the comparison:
\begin{itemize}
    \item \textbf{Contemporaneous ($c=0$):} $D_{\{i, 0\}}$ is the set of distances between paper $i$ and all other papers published up to and including its own publication year $t_i$.
    \item \textbf{Prospective ($c>0$):} $D_{\{i, 3\}}$ is the set of distances between paper $i$ and all papers published up to three years after its publication.
    \item \textbf{Retrospective ($c<0$):} $D_{\{i, -3\}}$ is the set of distances between paper $i$ and all papers that were published at least three years prior to its own publication.
\end{itemize}

\subsection{Point-in-Time Isolation Metrics}
The following four metrics quantify a paper's isolation at a specific moment in time, defined by the offset $c$. These metrics form the basis of our spatial novelty score. For clarity, a higher value for distance-based metrics indicates greater isolation, while a lower value for density-based metrics indicates greater isolation.

\paragraph{Metric 1: Neighborhood Density.}
The number of papers within a radius defined by a fraction $\gamma$ of the mean distance from paper $i$:
\[
N(i, c) = \big\lvert \{ d\in D_{\{ i, c \}} \mid d\leq \gamma\cdot Mean\left( D_{\{ i, c \}} \right) \} \big\rvert
\]

\paragraph{Metric 2: $k-$Nearest Neighbors ($k-$NN) Distance.} 
The distance to the $k-$th closest paper in the comparison set:
\[
d^k_{\{i, c\}} = k\text{-th smallest element of } D_{\{i, c\}}
\]

A larger value of $d^k_{\{i, c\}}$ indicates that a paper's closest intellectual peers are semantically distant, implying higher isolation.

\paragraph{Metric 3: Average $k-$NN  Distance.} The average distance to the $k$ nearest neighbors:
\[
\bar{d}^k_{\{i, c\}} = \frac{1}{k}\sum_{j=1}^k \left(j\text{-th smallest element of } D_{\{i, c\}}\right)
\]

This provides a more robust measure of local isolation than the single $k$-NN distance. A higher value indicates greater isolation.

\paragraph{Metric 4: Kernel Density.} A Gaussian kernel density estimate of the local semantic environment:
\[
density(i, c, k) = \sum_{d\in D_{\{i, c\}}} e^{-d^2/2\sigma^2}
\]

The bandwidth $\sigma$ is adapted for each paper by setting it to that paper's $k$-NN distance, $\sigma = d^k_{\{i, c\}}$. This provides a smooth, weighted measure of local density. 

\subsection{Dynamic Isolation Metrics}
Dynamic metrics capture temporal novelty by measuring the change in a paper's semantic environment over time. 
For any point-in-time metric $M$ defined above, we define its change between two time points, specified by offsets $c_1$ and $c_2$ where $c_1<c_2$, as:
\[
\Delta M(i, c_1, c_2) = M(i, c_1) - M(i, c_2)
\]

The interpretation of the sign of $\Delta M$ depends on whether $M$ is a distance or density metric. We primarily use two types of dynamic metrics:

\begin{itemize}
\item \textbf{Retrospective Change:} The change in isolation in the period leading up to a paper's publication. For example, using the $k$-NN distance metric, we can calculate $\Delta d^k(i, -3, 0) = d^k_{\{i, -3\}} - d^k_{\{i, 0\}}$. A large positive value for this measure indicates that the $k$-NN distance decreased significantly over the three years prior to publication, implying the neighborhood became rapidly denser. This signals that the paper is entering a "hot" or consolidating research frontier.
\item \textbf{Prospective Change:} The change in isolation in the period following the focal paper's publication, such as $\Delta d^k(i, 0, 3) = d^k_{\{i, 0\}} - d^k_{\{i, 3\}}$. While not used for prediction (as it uses future information), this serves as a validation measure. A large positive value indicates that a paper's neighborhood became much denser after it was published, suggesting it was a precursor to a new line of inquiry.
\end{itemize}

\subsection{Baseline Semantic Indicators}
\label{app-sec:baseline_semantic_metric}
To isolate paper-specific novelty from corpus-wide trends (e.g., the overall densification of the semantic space over time), we construct a set of baseline indicators. These serve as control variables for semantic isolation metrics.
\begin{itemize}
\item \textbf{Paper-level Mean Distance:} For each paper $i$, its mean distance to all prior work: $Mean (D_{ \{ i,0 \} })$. 
\item \textbf{Corpus-level Annual Mean Distance:} For each year $t$, the average distance between all pairs of papers published in or before that year. The measure is $Mean_{i,j\in S(t),i\neq j} \left( d(e_i,e_j) \right)$.
\item \textbf{Temporal Change in Corpus-level Distance:} The year-over-year change in the corpus-level annual mean distance. This controls for secular trends in the semantic space.
\end{itemize}


\section{Database Construction and Feature Implementation}
\label{app-sec:data}

This appendix provides a detailed description of the data collection, processing, and feature engineering pipeline used in our analysis.

\subsection{Dataset Construction}

\paragraph{Data Sources and Scope.}
Our dataset is constructed from three primary sources: RePEc (Research Papers in Economics) for metadata, Google Scholar for citation counts and PDF access, and SciSciNet \citep{lin2023sciscinet} for bibliometric indicators. The scope of our corpus is defined by the list of economics journals compiled by \citet{angrist2017economic}, which includes a broad set of general interest and top field journals.

\paragraph{Data Collection and Integration.}
The data construction process involved three main steps. First, we programmatically collected article metadata (title, author, year, journal, volume) from the RePEc website for all journals in our target list. Second, we queried each article title on Google Scholar to retrieve citation counts (as of December 2024) and links to publicly available PDF versions. Third, we downloaded the full-text PDFs and matched each article to its corresponding entry in the SciSciNet database using title, author, and publication year to retrieve pre-computed disruption and atypicality scores. We performed extensive cleaning to retain only main research articles and ensure accurate matching across sources.

\paragraph{Sample Definition.}
As described in Section \ref{sec:data}, the final dataset is partitioned to facilitate a causally-grounded analysis.
\begin{itemize}
\item \textbf{Knowledge Base (1980–1995):} This set of papers provides the historical semantic context against which the novelty of later papers is measured.
\item \textbf{Focal Period (1996–2015):} This is the primary analysis sample, for which we measure novelty and subsequent impact.
\end{itemize}
After filtering and matching procedures, the final analysis dataset contains 52,766 papers from the focal period with complete metadata, text, and impact data. A list of the journals included is provided in Table \ref{app-tab:econ_journals}.

\begin{table}[ht]
\centering
\tiny
\begin{tabular}{|l|l|}
\hline
\textbf{Journal Name} & \textbf{Journal Name} \\
\hline
American Economic Review & American Economist \\
American Journal of Agricultural Economics & Annals of Economic and Social Measurement \\
Applied Economics & Bell Journal of Economics \\
Bell Journal of Economics and Management Science & Brookings Papers on Economic Activity \\
Canadian Journal of Economics & Carnegie-Rochester Conference Series on Public Policy \\
Econometric Theory & Econometrica \\
Economic Development and Cultural Change & Economic Inquiry \\
Economic Record & Economic Theory \\
Economica & Economics Letters \\
European Economic Review & Explorations in Economic History \\
Federal Reserve Bank of St. Louis Review & Games and Economic Behavior \\
Industrial and Labor Relations Review & International Economic Review \\
International Journal of Game Theory & International Journal of Industrial Organization \\
International Labour Review & Journal of Business \& Economic Statistics \\
Journal of Econometrics & Journal of Economic Behavior \& Organization \\
Journal of Economic Dynamics and Control & Journal of Economic History \\
Journal of Economic Issues & Journal of Economic Literature \\
Journal of Economic Perspectives & Journal of Economic Theory \\
Journal of Economics \& Management Strategy & Journal of Environmental Economics and Management \\
Journal of Health Economics & Journal of Human Resources \\
Journal of Industrial Economics & Journal of Institutional and Theoretical Economics \\
Journal of International Economics & Journal of International Money and Finance \\
Journal of Labor Economics & Journal of Law and Economics \\
Journal of Law, Economics, \& Organization & Journal of Legal Studies \\
Journal of Mathematical Economics & Journal of Monetary Economics \\
Journal of Money, Credit and Banking & Journal of Political Economy \\
Journal of Productivity Analysis & Journal of Public Economics \\
Journal of Regional Science & Journal of Regulatory Economics \\
Journal of Urban Economics & Journal of the European Economic Association \\
Kyklos & NBER Macroeconomics Annual \\
National Tax Journal & Oxford Bulletin of Economics and Statistics \\
Oxford Economic Papers (New Series) & Public Policy \\
Quarterly Journal of Economics & Quarterly Review of Economics and Business \\
Quarterly Review of Economics and Finance & RAND Journal of Economics \\
Review of Economic Dynamics & Review of Economic Studies \\
Review of Economics and Statistics & Review of Income and Wealth \\
Review of Radical Political Economics & Review of Social Economy \\
Social Choice and Welfare & Southern Economic Journal \\
The Economic Journal & The Energy Journal \\
Theory and Decision & World Development \\
\hline
\end{tabular}
\caption{List of Economics Journals included in the Corpus. We use the list of journals from \cite{angrist2017economic}.}
\label{app-tab:econ_journals}
\end{table}

\subsection{Semantic Content Extraction Pipeline}
A central innovation of our study is the extraction of structured, multi-dimensional representations of each paper's intellectual contribution using Large Language Models (LLMs).
\paragraph{Text Processing.}
The pipeline begins with the raw PDF files. Each PDF is converted to plain text using an LLM-based Optical Character Recognition (OCR) model to handle complex layouts and mathematical notation. To manage computational costs while preserving essential content, we employ a two-step summarization process using Gemini Pro. First, the full text is condensed into an structured comprehensive summary (approx. 1500–2000 words). This summary is designed to retain all core arguments, methods, and findings.

\paragraph{Multi-Dimensional Insight Extraction.}
Second, from this comprehensive summary, we prompt the LLM to extract five distinct aspects of the paper's contribution. This decomposition allows us to measure novelty along different intellectual dimensions. The five insight types are:

\begin{itemize}
\item \textbf{Paragraph}: A concise overview of the paper's primary contribution.
\item \textbf{Topic}: The paper's subfield classification, research subject, and specific issues addressed
\item \textbf{Question-Conclusion}: The core research questions posed and the principal findings or conclusions.
\item \textbf{Methodology}: The data sources, empirical strategy, analytical models, and technical approaches used.
\item \textbf{Theory}: The theoretical frameworks, conceptual models, and analytical innovations introduced.
\end{itemize}

The prompts for insights and comprehensive summary are in Appendix \ref{app-sec:prompt}.

\subsection{Feature Engineering}
The final stage of our pipeline involves generating the semantic isolation metrics from the structured text data.
\paragraph{Embedding and Distance Calculation.}
To ensure the robustness of our semantic representations, we generate embeddings for each of the five insight types using three distinct models: SciBERT, SPECTER, and Qwen. We then calculate pairwise distances between all papers in the corpus using both Euclidean distance and cosine similarity.

\paragraph{Systematic Feature Construction.}
The full set of 3,240 candidate features is generated by systematically combining choices along five axes:

\begin{enumerate}
\item \textbf{Insight Type (5 options):} Paragraph, Topic, Question-Conclusion, Methodology, Theory.
\item \textbf{Embedding Model (3 options):} SciBERT, SPECTER, Qwen.
\item \textbf{Distance Metric (2 options):} Euclidean, Cosine similarity.
\item \textbf{Isolation Metric Type:} Neighborhood Density, $k$-NN Distance, Average $k$-NN Distance, Kernel Density, and the Dynamic metrics.
\item \textbf{Hyperparameters:} The metrics are calculated across a range of parameters to capture isolation at different scales:
\begin{itemize}
\item For Neighborhood Density: $\gamma\in \{0.5,1.0,1.5\}$.
\item For $k$-NN based metrics: $k\in \{ 3,5,10, 20, 30, 50\}$.
\item For Dynamic Change metrics: The change is computed over intervals defined by offsets $c_1, c_2\in \{-5,-3,0\}$ years relative to the publication date (e.g., $[-3, 0]$ for retrospective).
\end{itemize}
\end{enumerate}
This systematic, combinatorial approach ensures a comprehensive measurement of a paper's semantic position and trajectory, forming the empirical foundation for our analysis.

\section{More Results on Validating Semantic Isolation Metrics}
\label{app-sec:more_res_predictive}

This appendix provides supplementary tables and analyses that support the validation of our semantic isolation metrics presented in Section \ref{sec:prediction}.

\subsection{Full Results for Predictive Tasks}
\label{app-sec:more_res_full}
This section presents the complete set of results for the prediction tasks, comparing the performance of our Isolation metrics against all other feature groups.

\paragraph{Predicting Log Citation Counts.}
Table \ref{app-tab:log_cite_individual} details the performance of each individual feature group in predicting log citation counts. Notably, the Isolation feature set demonstrates predictive power (measured by MSE reduction) that is highly competitive with standard textual representations like TF-IDF. Table \ref{app-tab:log_cite_add_to_all} shows that the dimensionally-reduced Isolation-PCA feature set (200 components) retains nearly all the predictive power of the full set, justifying its use in the SHAP analysis in the main text and confirming that the signal is robust, not an artifact of high dimensionality.

\begin{table}[ht]
    \centering
\begin{tabular}{lll}
\toprule
feature set & MSE & comp \\
\midrule
control & 1.7529 & -0.0285 \\
 & (0.00161) &  \\
control+subfield & 1.7044 & 0.0 \\
 & (0.00154) &  \\
control+journal & 1.3252 & 0.2225 \\
 & (0.00122) &  \\
control+textual & 1.4558 & 0.1459 \\
 & (0.00139) &  \\
control+LDA & 1.5698 & 0.0790 \\
 & (0.00149) &  \\
control+TFIDF & 1.3935 & 0.1824 \\
 & (0.00135) &  \\
control+atypicality & 1.7110 & -0.0039 \\
 & (0.00156) &  \\
control+isolation & 1.5352 & 0.0993 \\
 & (0.00133) &  \\
control+isolation-PCA & 1.5728 & 0.0772 \\
 & (0.00137) &  \\
\bottomrule
\end{tabular}
    \caption{Individual feature groups predicting log citation counts.}
    \label{app-tab:log_cite_individual}
\end{table}

\begin{table}[ht]
    \centering
\begin{tabular}{lll}
\toprule
feature set & MSE & comp \\
\midrule
control+subfield & 1.7044 & 0.0 \\
 & (0.00154) &  \\
all features (excl. isolation) & 1.1465 & 0.3273 \\
 & (0.00111) &  \\
all features (incl. isolation) & 1.1092 & 0.3492 \\
 & (0.00105) &  \\
all features (incl. isolation-PCA) & 1.1105 & 0.3485 \\
 & (0.00105) &  \\
\bottomrule
\end{tabular}
    \caption{Additional value of isolation-PCA in predicting log citations.}
    \label{app-tab:log_cite_add_to_all}
\end{table}

\paragraph{Predicting Disruptive Papers.}
The full results for the disruption prediction task are presented in Tables \ref{app-tab:DI_paper_individual} and \ref{app-tab:DI_paper_add_to_all}. As with citations, the Isolation metrics are a top-performing feature group, with precision and F1-scores comparable to TF-IDF. The Isolation-PCA features again provide a significant, additional improvement in predictive performance, reinforcing the findings from the main text.

\begin{table}[ht]
    \centering
\begin{tabular}{lllllll}
\toprule
Feature set & Precision & Comp-prec & F1 & Comp-F1 \\
\midrule
control & 0.2875 & -0.0387 & 0.3885 & -0.0455 \\
 & (0.00046) &  & (0.00049) &  \\
control+subfield & 0.3140 & 0.0 & 0.4151 & 0.0 \\
 & (0.00053) &  & (0.00057) &  \\
control+journal & 0.3398 & 0.0377 & 0.4384 & 0.0397 \\
 & (0.00055) &  & (0.00054) &  \\
control+textual & 0.3786 & 0.0942 & 0.4597 & 0.0762 \\
 & (0.00068) &  & (0.00062) &  \\
control+LDA & 0.3717 & 0.0841 & 0.4634 & 0.0825 \\
 & (0.00059) &  & (0.00056) &  \\
control+TFIDF & 0.4171 & 0.1503 & 0.4835 & 0.1169 \\
 & (0.00067) &  & (0.00052) &  \\
control+atypical & 0.3231 & 0.0133 & 0.4220 & 0.0118 \\
 & (0.00057) &  & (0.00058) &  \\
control+isolation & 0.4064 & 0.1347 & 0.4699 & 0.0937 \\
 & (0.00064) &  & (0.00059) &  \\
control+isolation-PCA & 0.4007 & 0.1263 & 0.4599 & 0.0764 \\
 & (0.00074) &  & (0.00063) &  \\
\bottomrule
\end{tabular}
    \caption{Individual feature groups predicting disruptive papers.}
    \label{app-tab:DI_paper_individual}
\end{table}

\begin{table}[ht]
    \centering
\begin{tabular}{lllllll}
\toprule
Feature set & Precision & Comp-prec & F1 & Comp-F1 \\
\midrule
control & 0.2875 & -0.0387 & 0.3885 & -0.0455 \\
 & (0.00046) &  & (0.00049) &  \\
control+subfield & 0.3140 & 0.0 & 0.4151 & 0.0 \\
 & (0.00053) &  & (0.00057) &  \\
all & 0.4270 & 0.1647 & 0.4915 & 0.1305 \\
 & (0.0007) &  & (0.0006) &  \\
all+isolation & 0.4517 & 0.2007 & 0.5063 & 0.1559 \\
 & (0.00081) &  & (0.00067) &  \\
all+isolation-PCA & 0.4514 & 0.2003 & 0.5053 & 0.1541 \\
 & (0.00082) &  & (0.00066) &  \\
\bottomrule
\end{tabular}
    \caption{Additional value of isolation-PCA in predicting disruption.}
    \label{app-tab:DI_paper_add_to_all}
\end{table}

\clearpage

\subsection{Decomposition of Predictive Power by Insight Type}
\label{app-sec:isolation_insight}

To understand which dimensions of a paper's content contribute most to our novelty measures, we decompose the Isolation feature set by the five LLM-generated insight types. Tables \ref{app-tab:log_cite_insight} through \ref{app-tab:DI_paper_insight_add_to_all} present the results.
For both citations and disruption, isolation metrics based on the Paragraph Summary and Question-Conclusion consistently yield the highest predictive performance. This reinforces the finding that novelty in a paper's central claims is a primary predictor of impact. Metrics based on Theory also perform strongly. While Methodology and Topic isolation show slightly weaker performance, they still provide significant predictive value, demonstrating that novelty along these dimensions is also an important, albeit secondary, signal of future influence.

\begin{table}[ht]
    \centering
\begin{tabular}{lll}
\toprule
Feature set & MSE & Completeness \\
\midrule
control+subfield & 1.7044 & 0.0 \\
 & (0.00154) &  \\
control+atypical & 1.7110 & -0.0039 \\
 & (0.00156) &  \\
control+isolation & 1.5352 & 0.0993 \\
 & (0.00133) &  \\
control+isolation-paragraph & 1.5914 & 0.0663 \\
 & (0.00145) &  \\
control+isolation-methodology & 1.6191 & 0.0500 \\
 & (0.00137) &  \\
control+isolation-theory & 1.6182 & 0.0506 \\
 & (0.00142) &  \\
control+isolation-topic & 1.6342 & 0.0412 \\
 & (0.00146) &  \\
control+isolation-question-conclusion & 1.6060 & 0.0577 \\
 & (0.00139) &  \\
\bottomrule
\end{tabular}
    \caption{Predicting log citations: performance of individual insight types.}
    \label{app-tab:log_cite_insight}
\end{table}

\begin{table}[ht]
    \centering
\begin{tabular}{lll}
\toprule
Feature set & MSE & Completeness \\
\midrule
control+subfield & 1.7044 & 0.0 \\
 & (0.00154) &  \\
all features (excl. isolation) & 1.1465 & 0.3273 \\
 & (0.00111) &  \\
all features (incl. isolation) & 1.1092 & 0.3492 \\
 & (0.00105) &  \\
all features (incl. isolation-paragraph) & 1.1065 & 0.3508 \\
 & (0.00109) &  \\
all features (incl. isolation-methodology) & 1.1359 & 0.3335 \\
 & (0.00112) &  \\
all features (incl. isolation-theory) & 1.1175 & 0.3444 \\
 & (0.00107) &  \\
all features (incl. isolation-topic) & 1.1313 & 0.3363 \\
 & (0.00105) &  \\
all features (incl. isolation-ques-conc) & 1.1118 & 0.3477 \\
 & (0.00106) &  \\
\bottomrule
\end{tabular}
    \caption{Predicting log citations: additional value of insight types}
    \label{app-tab:log_cite_insight_add_to_all}
\end{table}

\begin{table}[ht]
    \centering
\begin{tabular}{lllllll}
\toprule
feature & prec & comp-prec & f1 & comp-f1 \\
\midrule
control & 0.2875 & -0.0387 & 0.3885 & -0.0455 \\
 & (0.00046) &  & (0.00049) &  \\
control+subfield & 0.3140 & 0.0 & 0.4151 & 0.0 \\
 & (0.00053) &  & (0.00057) &  \\
control+journal & 0.3398 & 0.0377 & 0.4384 & 0.0397 \\
 & (0.00055) &  & (0.00054) &  \\
control+textual & 0.3786 & 0.0942 & 0.4597 & 0.0762 \\
 & (0.00068) &  & (0.00062) &  \\
control+LDA & 0.3717 & 0.0841 & 0.4634 & 0.0825 \\
 & (0.00059) &  & (0.00056) &  \\
control+TFIDF & 0.4171 & 0.1503 & 0.4835 & 0.1169 \\
 & (0.00067) &  & (0.00052) &  \\
control+atypical & 0.3231 & 0.0133 & 0.4220 & 0.0118 \\
 & (0.00057) &  & (0.00058) &  \\
control+isolation & 0.4064 & 0.1347 & 0.4699 & 0.0937 \\
 & (0.00064) &  & (0.00059) &  \\
control+isolation-paragraph & 0.3773 & 0.0923 & 0.4556 & 0.0691 \\
 & (0.00067) &  & (0.00063) &  \\
control+isolation-methodology & 0.3618 & 0.0697 & 0.4292 & 0.0241 \\
 & (0.00071) &  & (0.00065) &  \\
control+isolation-theory & 0.3610 & 0.0685 & 0.4360 & 0.0356 \\
 & (0.00059) &  & (0.00051) &  \\
control+isolation-topic & 0.3505 & 0.0533 & 0.4191 & 0.0068 \\
 & (0.00065) &  & (0.00055) &  \\
control+isolation-question-conclusion & 0.3662 & 0.0762 & 0.4416 & 0.0453 \\
 & (0.00058) &  & (0.00056) &  \\
\bottomrule
\end{tabular}
    \caption{Predicting disruption: performance of individual insight types}
    \label{app-tab:DI_paper_insight}
\end{table}

\begin{table}[ht]
    \centering
\begin{tabular}{lllllll}
\toprule
Feature set & Precision & Comp-prec & F1 & Comp-F1 \\
\midrule
control & 0.2875 & -0.0387 & 0.3885 & -0.0455 \\
 & (0.00461) &  & (0.00486) &  \\
control+subfield & 0.3140 & 0.0 & 0.4151 & 0.0 \\
 & (0.0053) &  & (0.0057) &  \\
all features (excl. isolation) & 0.4270 & 0.1647 & 0.4915 & 0.1305 \\
 & (0.00701) &  & (0.00602) &  \\
all features (incl. isolation) & 0.4517 & 0.2007 & 0.5063 & 0.1559 \\
 & (0.00081) &  & (0.00067) &  \\
all features (incl. isolation-paragraph) & 0.4419 & 0.1864 & 0.5062 & 0.1557 \\
 & (0.00073) &  & (0.00062) &  \\
all features (incl. isolation-method) & 0.4338 & 0.1746 & 0.4932 & 0.1334 \\
 & (0.00072) &  & (0.0006) &  \\
all features (incl. isolation-theory) & 0.4380 & 0.1808 & 0.5003 & 0.1457 \\
 & (0.00073) &  & (0.00063) &  \\
all features (incl. isolation-topic) & 0.4357 & 0.1774 & 0.4961 & 0.1385 \\
 & (0.00075) &  & (0.00058) &  \\
all features (incl. isolation-ques-conc) & 0.4404 & 0.1843 & 0.5029 & 0.1500 \\
 & (0.00073) &  & (0.00061) &  \\
\bottomrule
\end{tabular}
    \caption{predicting disruption: additional value of insight types.}
    \label{app-tab:DI_paper_insight_add_to_all}
\end{table}

\clearpage

\subsection{Subgroup Analysis by Economic Subfield}
\label{app-sec:isolation_specific_group}

To assess the generalizability of our findings, we conduct the prediction exercises separately for each of the 16 economic subfields in our sample. Tables \ref{app-tab:log_cite_subfield} and \ref{app-tab:DI_paper_subfield} report the results for predicting log citations and disruption, respectively.

While the overall predictive power of isolation metrics holds across most fields, the relative importance of different insight types varies. Notably, isolation metrics derived from the Question-Conclusion summary are the strongest individual predictors in a majority of subfields for both citations and disruption. This suggests that novelty in a paper's core argument - the questions it asks and the answers it provides - is a particularly potent and generalizable predictor of scientific impact. We also observe that overall predictive accuracy tends to be lower in more established or theoretical fields (e.g., Microeconomic Theory, Macroeconomics), which also exhibit lower baseline rates of disruption.

\begin{table}[ht]
    \centering
    \scriptsize
    \rotatebox{90}{  
\begin{tabular}{llllllll}
\toprule
field & control & +atypical & +iso & +method & +theory & +topic & +qu-co \\
\midrule
macroeconomics & 2.2014 (0.000) & 2.1250 (0.035) & 1.8342 (0.167) & 1.9457 (0.116) & 1.9157 (0.130) & 1.9562 (0.111) & 1.9141 (0.131) \\
 & (0.00062) & (0.00065) & (0.00058) & (0.00055) & (0.00060) & (0.00057) & (0.00058) \\
international & 2.1205 (0.000) & 2.0368 (0.040) & 1.7264 (0.186) & 1.8271 (0.138) & 1.8125 (0.145) & 1.8404 (0.132) & 1.7916 (0.155) \\
 & (0.00065) & (0.00064) & (0.00055) & (0.00063) & (0.00059) & (0.00057) & (0.00058) \\
financial & 2.1997 (0.000) & 2.0934 (0.048) & 1.8225 (0.171) & 1.9334 (0.121) & 1.8914 (0.140) & 1.9251 (0.125) & 1.8678 (0.151) \\
 & (0.00066) & (0.00064) & (0.00056) & (0.00061) & (0.00057) & (0.00063) & (0.00054) \\
economic history & 1.6489 (0.000) & 1.5483 (0.061) & 1.3266 (0.195) & 1.3932 (0.155) & 1.3978 (0.152) & 1.3635 (0.173) & 1.3631 (0.173) \\
 & (0.00103) & (0.00099) & (0.00106) & (0.00093) & (0.00093) & (0.00099) & (0.00099) \\
microeconomics theory & 1.7256 (0.000) & 1.6635 (0.036) & 1.4741 (0.146) & 1.5519 (0.101) & 1.5235 (0.117) & 1.5330 (0.112) & 1.5263 (0.116) \\
 & (0.00040) & (0.00037) & (0.00036) & (0.00035) & (0.00038) & (0.00036) & (0.00035) \\
industrial organization & 1.7983 (0.000) & 1.7007 (0.054) & 1.5167 (0.157) & 1.5792 (0.122) & 1.5727 (0.125) & 1.6107 (0.104) & 1.5620 (0.131) \\
 & (0.00058) & (0.00050) & (0.00048) & (0.00051) & (0.00051) & (0.00051) & (0.00057) \\
econometrics \& quantitative & 2.1885 (0.000) & 2.1258 (0.029) & 1.7725 (0.190) & 1.8670 (0.147) & 1.8526 (0.153) & 1.8881 (0.137) & 1.8547 (0.153) \\
 & (0.00059) & (0.00060) & (0.00049) & (0.00054) & (0.00049) & (0.00050) & (0.00049) \\
public & 1.7914 (0.000) & 1.7073 (0.047) & 1.5004 (0.162) & 1.5653 (0.126) & 1.5647 (0.127) & 1.5670 (0.125) & 1.5614 (0.128) \\
 & (0.00051) & (0.00060) & (0.00046) & (0.00051) & (0.00043) & (0.00049) & (0.00047) \\
health & 1.6354 (0.000) & 1.5511 (0.052) & 1.3217 (0.192) & 1.3811 (0.156) & 1.3758 (0.159) & 1.4088 (0.139) & 1.3791 (0.157) \\
 & (0.00081) & (0.00068) & (0.00064) & (0.00069) & (0.00068) & (0.00068) & (0.00061) \\
development & 1.8143 (0.000) & 1.7386 (0.042) & 1.4770 (0.186) & 1.5588 (0.141) & 1.5665 (0.137) & 1.5652 (0.137) & 1.5172 (0.164) \\
 & (0.00083) & (0.00073) & (0.00069) & (0.00066) & (0.00070) & (0.00069) & (0.00062) \\
labor & 1.9207 (0.000) & 1.8229 (0.051) & 1.5170 (0.210) & 1.6266 (0.153) & 1.6060 (0.164) & 1.6290 (0.152) & 1.5770 (0.179) \\
 & (0.00057) & (0.00049) & (0.00043) & (0.00045) & (0.00045) & (0.00046) & (0.00047) \\
agricultural & 1.5922 (0.000) & 1.4544 (0.087) & 1.2468 (0.217) & 1.3216 (0.170) & 1.2831 (0.194) & 1.3234 (0.169) & 1.2595 (0.209) \\
 & (0.00097) & (0.00086) & (0.00091) & (0.00091) & (0.00088) & (0.00091) & (0.00093) \\
urban \& regional & 1.8135 (0.000) & 1.6648 (0.082) & 1.4731 (0.188) & 1.5187 (0.163) & 1.5601 (0.140) & 1.5722 (0.133) & 1.5344 (0.154) \\
 & (0.00112) & (0.00109) & (0.00098) & (0.00104) & (0.00105) & (0.00090) & (0.00097) \\
environmental \& resource & 1.7978 (0.000) & 1.6711 (0.070) & 1.4622 (0.187) & 1.5374 (0.145) & 1.5127 (0.159) & 1.5355 (0.146) & 1.4934 (0.169) \\
 & (0.00098) & (0.00088) & (0.00097) & (0.00093) & (0.00093) & (0.00094) & (0.00093) \\
behavioral & 2.1318 (0.000) & 1.9431 (0.088) & 1.6765 (0.214) & 1.7759 (0.167) & 1.7541 (0.177) & 1.7583 (0.175) & 1.7319 (0.188) \\
 & (0.00076) & (0.00068) & (0.00065) & (0.00069) & (0.00070) & (0.00060) & (0.00066) \\
other & 2.2479 (0.000) & 2.0730 (0.078) & 1.9137 (0.149) & 2.0001 (0.110) & 1.9539 (0.131) & 1.9471 (0.134) & 1.9072 (0.152) \\
 & (0.00125) & (0.00103) & (0.00126) & (0.00125) & (0.00112) & (0.00118) & (0.00104) \\
\bottomrule
\end{tabular}
}
    \caption{The bootstrapped MSE of different feature groups when predicting log citation count of papers in different subfields. The value in parenthesis below the MSE is standard error of the 100 time bootstrapping, and the value in parenthesis on the right of the MSE is the completeness, setting control group as worst case (same to the subfield baseline) and perfect prediction as best case.}
    \label{app-tab:log_cite_subfield}
\end{table}

\begin{table}[ht]
    \centering
    \scriptsize
    \rotatebox{90}{  
\begin{tabular}{llllllll}
\toprule
field & control & +atypical & +iso & +method & +theory & +topic & +qu-co \\
\midrule
macroeconomics & 0.2947 (-0.000) & 0.3378 (0.061) & 0.4840 (0.268) & 0.4213 (0.179) & 0.3934 (0.140) & 0.4108 (0.165) & 0.4705 (0.249) \\
 & (0.00023) & (0.00028) & (0.00053) & (0.00046) & (0.00050) & (0.00064) & (0.00049) \\
international & 0.3464 (-0.000) & 0.4039 (0.088) & 0.5302 (0.281) & 0.4185 (0.110) & 0.4520 (0.162) & 0.4147 (0.105) & 0.4728 (0.193) \\
 & (0.00024) & (0.00025) & (0.00045) & (0.00038) & (0.00032) & (0.00039) & (0.00040) \\
financial & 0.2446 (-0.000) & 0.2874 (0.057) & 0.3815 (0.181) & 0.3561 (0.148) & 0.3535 (0.144) & 0.2927 (0.064) & 0.3224 (0.103) \\
 & (0.00033) & (0.00034) & (0.00081) & (0.00078) & (0.00066) & (0.00075) & (0.00073) \\
economic history & 0.3672 (-0.000) & 0.4097 (0.067) & 0.5280 (0.254) & 0.5060 (0.219) & 0.4981 (0.207) & 0.5049 (0.218) & 0.5323 (0.261) \\
 & (0.00051) & (0.00049) & (0.00065) & (0.00064) & (0.00067) & (0.00073) & (0.00067) \\
microeconomics theory & 0.1616 (-0.000) & 0.1863 (0.030) & 0.3701 (0.249) & 0.3181 (0.187) & 0.3027 (0.168) & 0.2944 (0.158) & 0.3005 (0.166) \\
 & (0.00017) & (0.00021) & (0.00060) & (0.00053) & (0.00046) & (0.00068) & (0.00051) \\
industrial organization & 0.3036 (-0.000) & 0.3566 (0.076) & 0.5147 (0.303) & 0.4252 (0.175) & 0.4831 (0.258) & 0.4404 (0.196) & 0.4655 (0.233) \\
 & (0.00024) & (0.00027) & (0.00049) & (0.00042) & (0.00047) & (0.00045) & (0.00040) \\
econometrics \& quantitative & 0.2535 (-0.000) & 0.3042 (0.068) & 0.4342 (0.242) & 0.3552 (0.136) & 0.3485 (0.127) & 0.3180 (0.086) & 0.3777 (0.166) \\
 & (0.00017) & (0.00022) & (0.00050) & (0.00041) & (0.00044) & (0.00037) & (0.00041) \\
public & 0.3531 (-0.000) & 0.4332 (0.124) & 0.5290 (0.272) & 0.4821 (0.199) & 0.4560 (0.159) & 0.4508 (0.151) & 0.4596 (0.165) \\
 & (0.00022) & (0.00028) & (0.00035) & (0.00037) & (0.00036) & (0.00036) & (0.00036) \\
health & 0.3415 (-0.000) & 0.3711 (0.045) & 0.5249 (0.279) & 0.4406 (0.151) & 0.4681 (0.192) & 0.4233 (0.124) & 0.4499 (0.165) \\
 & (0.00033) & (0.00038) & (0.00045) & (0.00053) & (0.00052) & (0.00045) & (0.00038) \\
development & 0.4689 (-0.000) & 0.5594 (0.170) & 0.6098 (0.265) & 0.5783 (0.206) & 0.5539 (0.160) & 0.5704 (0.191) & 0.5764 (0.202) \\
 & (0.00025) & (0.00030) & (0.00030) & (0.00027) & (0.00029) & (0.00031) & (0.00027) \\
labor & 0.3503 (-0.000) & 0.4180 (0.104) & 0.5145 (0.253) & 0.4614 (0.171) & 0.4570 (0.164) & 0.4642 (0.175) & 0.4746 (0.191) \\
 & (0.00018) & (0.00020) & (0.00032) & (0.00030) & (0.00025) & (0.00029) & (0.00032) \\
agricultural & 0.4809 (-0.000) & 0.5426 (0.119) & 0.6036 (0.236) & 0.5717 (0.175) & 0.5699 (0.171) & 0.5737 (0.179) & 0.5950 (0.220) \\
 & (0.00042) & (0.00046) & (0.00045) & (0.00042) & (0.00047) & (0.00046) & (0.00050) \\
urban \& regional & 0.3392 (-0.000) & 0.3531 (0.021) & 0.4983 (0.241) & 0.3997 (0.092) & 0.4756 (0.206) & 0.4446 (0.160) & 0.4301 (0.137) \\
 & (0.00056) & (0.00062) & (0.00092) & (0.00090) & (0.00075) & (0.00071) & (0.00083) \\
environmental \& resource & 0.4153 (-0.000) & 0.4617 (0.079) & 0.5594 (0.246) & 0.5387 (0.211) & 0.5066 (0.156) & 0.4891 (0.126) & 0.5161 (0.172) \\
 & (0.00038) & (0.00043) & (0.00053) & (0.00053) & (0.00054) & (0.00049) & (0.00051) \\
behavioral & 0.2504 (-0.000) & 0.3453 (0.127) & 0.3932 (0.190) & 0.4118 (0.215) & 0.3950 (0.193) & 0.2874 (0.049) & 0.3453 (0.127) \\
 & (0.00052) & (0.00058) & (0.00122) & (0.00112) & (0.00124) & (0.00110) & (0.00096) \\
other & 0.4122 (-0.000) & 0.4810 (0.117) & 0.5098 (0.166) & 0.4774 (0.111) & 0.4932 (0.138) & 0.4632 (0.087) & 0.4974 (0.145) \\
 & (0.00039) & (0.00045) & (0.00058) & (0.00050) & (0.00053) & (0.00055) & (0.00051) \\
\bottomrule
\end{tabular}
}
    \caption{The bootstrapped precision of different feature groups when predicting disruptive papers in different subfields. The value in parenthesis below the MSE is standard error of the 100 time bootstrapping, and the value in parenthesis on the right of the MSE is the completeness, setting control group as worst case (same to the subfield baseline) and perfect prediction as best case.}
    \label{app-tab:DI_paper_subfield}
\end{table}

\clearpage

\subsection{Including Prospective Features}
\label{app-sec:temporal_isolation}

In this section we add perspective dynamic isolation metrics and check the predicting performance. Specifically, for dynamic metrics: The change is computed over intervals defined by offsets $c_1, c_2\in \{-5,-3,0,3,5\}$ years relative to the publication date (e.g., $[-3, 0]$ for retrospective, $[0, 3]$ for prospective). We call this new set of isolation features \textit{isolation-wf}, meaning it is with future information.

\paragraph{Predicting Log Citation Counts}.
Table \ref{app-tab:log_cite_individual_perspective} presents the standalone predictive power of the atypicality and isolation feature groups. When added to the control variables, the atypicality score offers no improvement over the baseline. In stark contrast, our isolation metrics achieve a completeness score of 17.9\%, demonstrating substantial predictive power.

\begin{table}[ht]
    \centering
\begin{tabular}{lll}
\toprule
Feature set & MSE & Completeness \\
\midrule
control+subfield & 1.7044 & 0.0 \\
 & (0.00154) &  \\
control+atypicality & 1.7110 & -0.0039 \\
 & (0.00156) &  \\
control+isolation-wf & 1.3999 & 0.1787 \\
 & (0.00133) &  \\
\bottomrule
\end{tabular}
    \caption{Predicting log citation counts: individual feature groups. Standard errors from 100 bootstrap iterations in parentheses. Completeness is the proportional reduction in MSE relative to the baseline.}
    \label{app-tab:log_cite_individual_perspective}
\end{table}

\begin{table}[ht]
    \centering
\begin{tabular}{lll}
\toprule
Feature set & MSE & Completeness \\
\midrule
control+subfield & 1.7044 & 0.0 \\
 & (0.00154) &  \\
control+subfield+atypicality & 1.6646 & 0.0234 \\
 & (0.00154) &  \\
control+subfield+atypicality+isolation-wf & 1.3801 & 0.1903 \\
 & (0.00131) &  \\
\midrule
all features (excl. isolation) & 1.1465 & 0.3273 \\
 & (0.00111) &  \\
all features (incl. isolation) & 1.0439 & 0.3875 \\
 & (0.00102) &  \\
\bottomrule
\end{tabular}
    \caption{Predicting log citation counts: additional value of isolation metrics.}
    \label{app-tab:log_cite_add_to_atypicality_perspective}
\end{table}

Table \ref{app-tab:log_cite_add_to_atypicality_perspective} shows that adding atypicality to the baseline provides a modest 2.3\% improvement in completeness. However, subsequently adding our isolation metrics increases completeness to 19.0\% - an incremental gain of 16.7 percentage points. This strong additional effect provides compelling evidence that isolation and atypicality are complementary, measuring distinct aspects of novelty. Atypicality captures the novelty of a paper's inputs (unusual combinations of references), while our metrics capture the novelty of its output (the semantic content of its contribution).

The bottom panel of Table \ref{app-tab:log_cite_add_to_atypicality_perspective} shows that even when added to a model containing all other feature groups, our isolation metrics provide a statistically and economically significant improvement, increasing completeness by an additional 6.1 percentage points.

\begin{figure}[ht]
    \centering
    \includegraphics[width=0.5\linewidth]{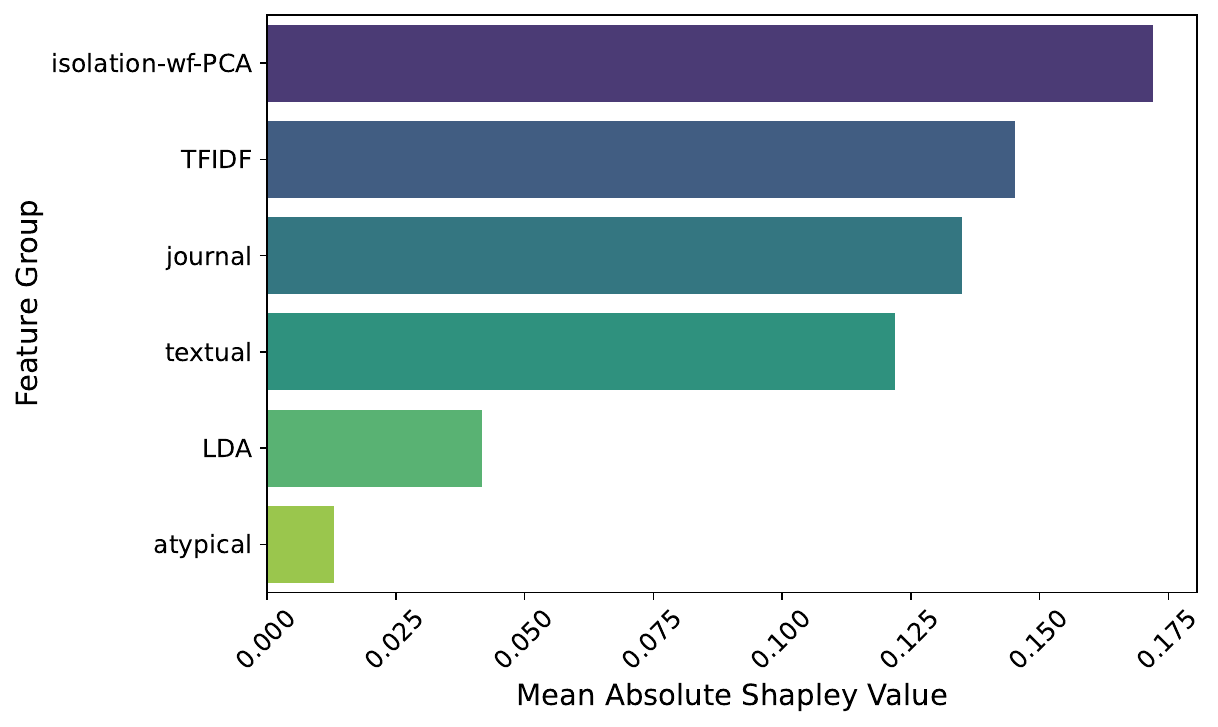}
    \caption{Feature importance (mean absolute SHAP) for predicting log citations.}
    \label{app-fig:log_cite_shap_cv_perspective}
\end{figure}

To assess the relative importance of each feature group, we compute SHAP (SHapley Additive exPlanations) values\footnote{We apply 5-fold cross validation, and combine the prediction on test folds to calculate the shap contribution.}. As shown in Figure \ref{app-fig:log_cite_shap_cv_perspective}, the isolation metrics (reduced to 200 dimensions via PCA for comparability) contribute more to the model's predictions than any other feature group, including TFIDF features. This confirms that semantic isolation is not a peripheral signal but a central predictor of scientific impact.

\paragraph{Predicting Disruptive Papers}

\begin{table}[ht]
    \centering
\begin{tabular}{lllll}
\toprule
Feature set & Precision & Comp-prec & F1 & Comp-F1 \\
\midrule
control+subfield & 0.3140 & 0.0 & 0.4151 & 0.0 \\
 & (0.00053) &  & (0.00057) &  \\
control+atypicality & 0.3231 & 0.0133 & 0.4220 & 0.0118 \\
 & (0.00057) &  & (0.00058) &  \\
control+isolation & 0.4115 & 0.1421 & 0.4692 & 0.0925 \\
 & (0.00072) &  & (0.00059) &  \\
\bottomrule
\end{tabular}
    \caption{Predicting disruptive papers: individual feature groups. Standard errors from 100 bootstrap iterations in parentheses. Comp-prec shows the completeness of error in precision.}
    \label{app-tab:DI_paper_individual_perspective}
\end{table}

The results for disruption prediction, presented in Table \ref{app-tab:DI_paper_individual_perspective}, mirror the patterns observed for citations. Our isolation metrics substantially outperform the atypicality score, achieving a 14.21\% completeness score on precision compared to just 1.33\% for atypicality. This consistency across different prediction tasks strengthens the evidence that isolation captures fundamental aspects of research novelty.

\begin{table}[ht]
    \centering
\begin{tabular}{lllll}
\toprule
Feature set & Precision & Comp-prec & F1 & Comp-F1 \\
\midrule
control+subfield & 0.3140 & 0.0 & 0.4151 & 0.0 \\
 & (0.00053) &  & (0.00057) &  \\
control+subfield+atypicality & 0.3437 & 0.0434 & 0.4424 & 0.0466 \\
 & (0.00054) &  & (0.00051) &  \\
control+subfield+atypicality+isolation & 0.4242 & 0.1606 & 0.4797 & 0.1104 \\
 & (0.00074) &  & (0.00062) &  \\
\midrule
all features (excl. isolation) & 0.4270 & 0.1647 & 0.4915 & 0.1305 \\
 & (0.00070) &  & (0.00060) &  \\
all features (incl. isolation) & 0.4551 & 0.2057 & 0.5052 & 0.1539 \\
 & (0.00075) &  & (0.00059) &  \\
\bottomrule
\end{tabular}
    \caption{Predicting disruptive papers: additional value of isolation metrics.}
    \label{app-tab:DI_paper_add_to_atypicality_perspective}
\end{table}

Moreover, the metrics are again complementary, as shown in Table~\ref{app-tab:DI_paper_add_to_atypicality_perspective}. Adding isolation metrics to a model that already includes the baseline and atypicality features increases precision completeness from 4.34\% to 16.06\%. This additional effect confirms that our measure of semantic distance captures a dimension of novelty relevant for disruption that is distinct from the recombinative novelty captured by atypicality.

The bottom panel of Table \ref{app-tab:DI_paper_add_to_atypicality_perspective} shows that isolation metrics continue to provide significant predictive value even in the presence of all other features, increasing precision completeness from 16.47\% to 20.57\%. 
Although the 4.1 percentage point improvement is more modest than for citation prediction, the standard error shows that the improvement in precision is still significant. It demonstrates the consistent and non-redundant value of isolation across different impact dimensions.

\begin{figure}[ht]
    \centering
    \includegraphics[width=0.5\linewidth]{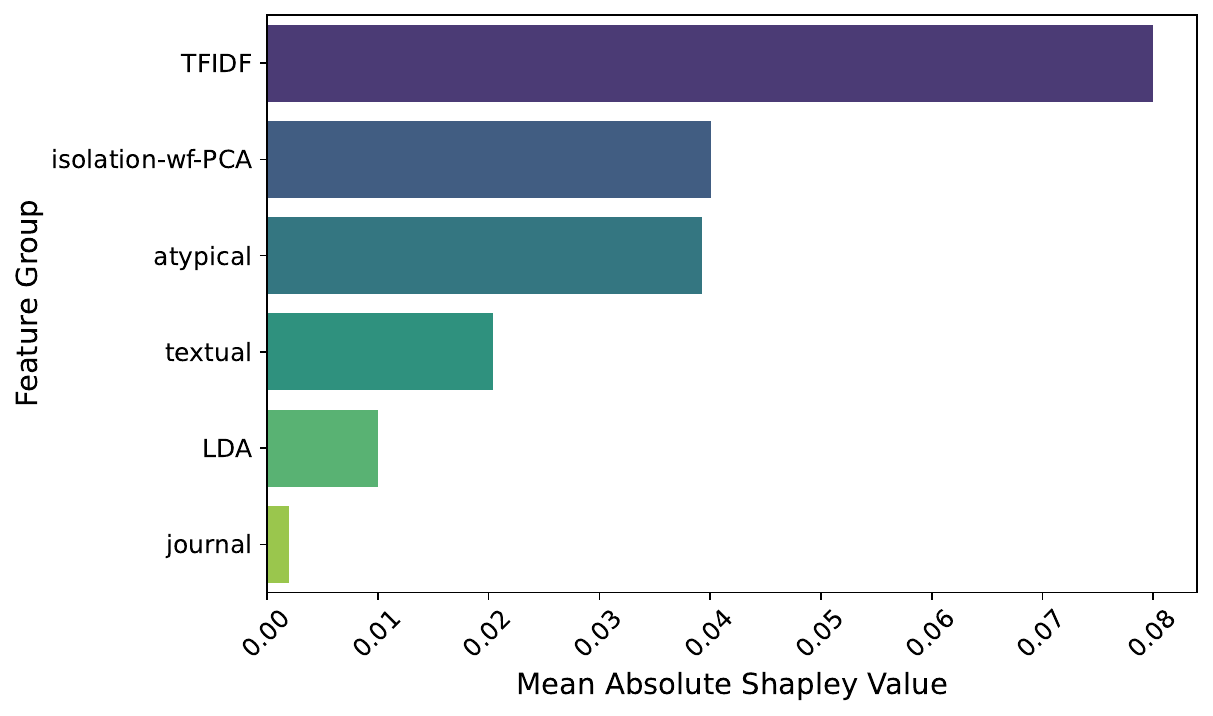}
    \caption{Feature importance (mean absolute SHAP) for predicting disruption.}
    \label{app-fig:DI_paper_shap_cv_perspective}
\end{figure}

The SHAP analysis\footnote{We apply 5-fold cross validation, and combine the prediction on test folds to calculate the shap contribution.} in Figure \ref{app-fig:DI_paper_shap_cv_perspective} further confirms that our isolation features are a key driver of the model's predictions.
This confirms that isolation captures meaningful signals for identifying potentially disruptive research, further validating our approach across different dimensions of research impact.

\clearpage

\section{More results on Typology}
\label{app-sec:quadrant_data_statistics}

This appendix provides technical details on the construction of our novelty scores and presents supplementary statistics on the distribution of papers within our two-dimensional framework.

\subsection{From Isolation Metrics to Novelty Scores via PCA}
\label{app-sec:PCA}
We construct the spatial and temporal novelty scores by applying Principal Component Analysis (PCA) to the relevant subsets of our standardized semantic isolation metrics. The first principal component (PC1) of each set serves as our composite index.
\begin{itemize}
\item \textbf{Spatial Novelty Score:} PC1 of the standardized contemporaneous point-in-time isolation metrics and the baseline paper-level mean distance.
\item \textbf{Temporal Novelty Score:} PC1 of the standardized retrospective dynamic isolation metrics.
\end{itemize}

\begin{figure}[ht!]
    \centering
    \begin{subfigure}[t]{0.48\linewidth}
        \centering
        \includegraphics[width=\linewidth]{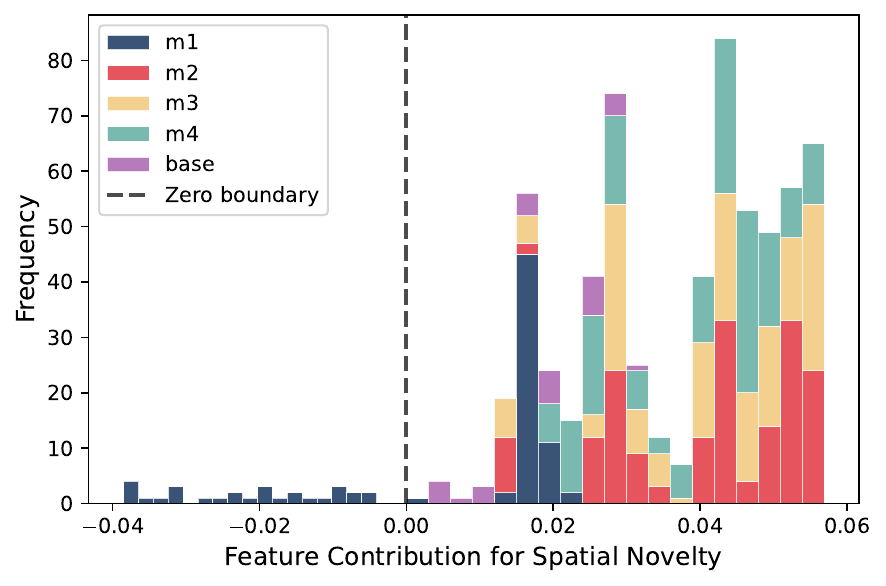}
        \caption{}
        \label{app-fig:pca_spatial}
    \end{subfigure}
    \hfill
    \begin{subfigure}[t]{0.48\linewidth}
        \centering
        \includegraphics[width=\linewidth]{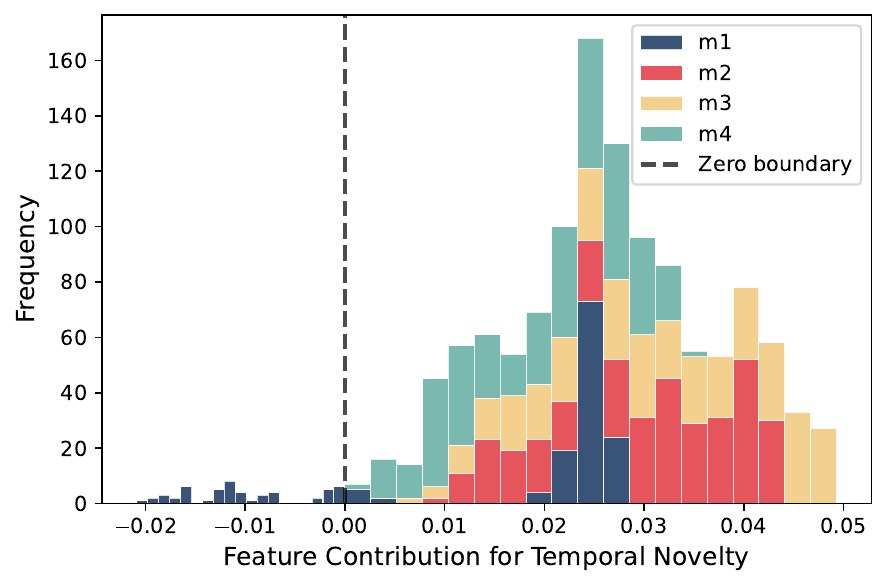}
        \caption{}
        \label{app-fig:pca_temporal}
    \end{subfigure}
    \caption{Distribution of feature contribution. Panel (a) is for spatial novelty, and panel (b) is for temporal novelty. 
    For the metric name: m1 is neighborhood density; m2 is $k-$NN distance; m3 is $k-$NN average distance; and m4 is Gaussian density.}
    \label{app-fig:pca_dist}
\end{figure}

To interpret the novelty scores, we examine the contributions (factor loadings) of the underlying features to the first principal component (PC1) for both the spatial and temporal analyses, as shown in Figure \ref{app-fig:pca_dist}.

For both novelty scores, the interpretation is anchored by the dominant, positive loadings of the distance-based metrics: 
$k$-NN distance (m2) and average $k$-NN distance (m3). This provides a clear and intuitive orientation for the final index. For the spatial novelty score, the strong positive contributions from m2 and m3 confirm that a higher value on PC1 corresponds to greater semantic isolation. For the temporal novelty score, their positive contributions confirm that a higher value on PC1 corresponds to a more rapidly densifying neighborhood (i.e., a larger positive change in isolation from $t=3$ to $t=0$), which is our measure of engagement with a "hot" research frontier.

The more complex loading patterns for the density-based metrics, particularly the mixed signs for neighborhood density (m1), reflect the imperfect correlations between the different proxies for novelty. While all four metric types are designed to capture aspects of novelty, they are not perfectly collinear. PCA constructs the optimal linear combination that captures the maximum shared variance. In this process, a feature might receive a small or even negative loading if doing so helps to "purify" the principal component, making it a better index of the primary variation defined by the dominant m2 and m3 metrics. This is a standard outcome of PCA when synthesizing multiple, related indicators.

Therefore, the resulting PC1 is a statistically robust index that synthesizes information from all our isolation measures, with its overall direction and interpretation reliably governed by the strongest and most intuitive distance-based features.

An analysis of the top-20 most influential features for each score reveals that a diverse set of metrics contributes to the final indices. For both scores, metrics derived from various insight types (especially Question-Conclusion and Paragraph Summary), embedding models (SPECTER and Qwen), and distance measures are all highly represented. This diversity underscores the value of our comprehensive feature engineering approach and demonstrates that the resulting novelty scores are robust composites rather than being driven by a single measurement choice.


\subsection{Descriptive Statistics of the Novelty Framework}
\label{app-sec:descriptive_stat}
\begin{figure}[h!]
    \centering
    \includegraphics[width=0.5\linewidth]{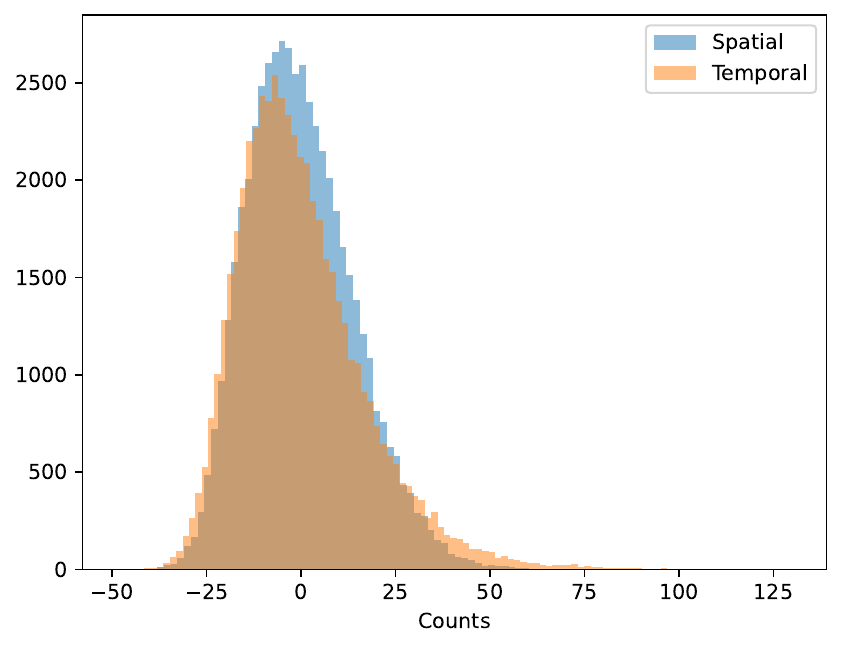}
    \caption{Distribution of novelty scores.}
    \label{app-fig:novelty_score_dist}
\end{figure}

Figure \ref{app-fig:novelty_score_dist} shows the distribution of spatial novelty score and temporal novelty score.

Table \ref{app-tab:quadrant_stat_count} presents the distribution of papers across the four novelty archetypes, defined by the median splits of the two scores. The prevalence of the off-diagonal archetypes - Trendy and Outlying neighborhoods - quantifies the structural trade-off discussed in the main text. Trailblazing neighborhoods, where papers successfully combine high spatial and temporal novelty, represent the least common archetype, comprising about one-fifth of the sample.

\begin{table}[h!]
    \centering
    \begin{tabular}{c|c|c}
    & & \\
    High Temporal          & Trendy & Trailblazing \\
                        & (N=15468, Percent=29.3\%) & (N=10915, Percent=20.7\%) \\
    \hline
    & & \\
    Low Temporal           & Consolidating & Outlying \\
                     & (N=10915, Percent=20.7\%) & (N=15468, Percent=29.3\%) \\
    \hline
    &  &  \\
         & Low Spatial & High Spatial \\
    \end{tabular}
    \caption{Distribution of papers across the four novelty archetypes.}
    \label{app-tab:quadrant_stat_count}
\end{table}

\begin{figure}
    \centering
    \includegraphics[width=0.55\linewidth]{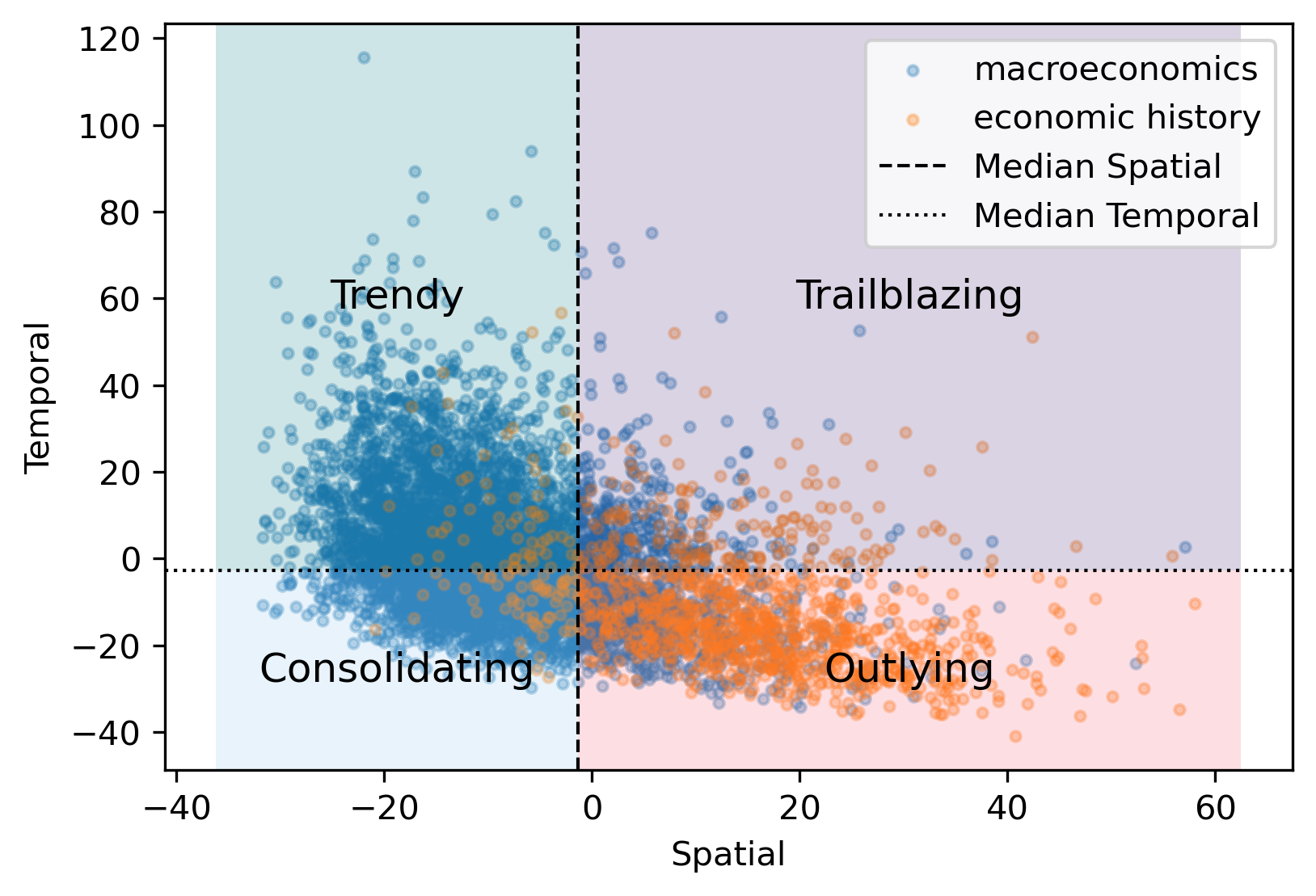}
    \caption{Distribution of macroeconomics and economic history papers, illustrating the distinct central tendencies but substantial overlap between the two fields.}
    \label{app-fig:macro_hist_scatter}
\end{figure}

Table \ref{app-tab:quadrant_stat_field} details the distribution of papers from each economic subfield across the four quadrants, reinforcing the field-mapping analysis in the main text. For example, Macroeconomics papers are heavily concentrated in the Consolidating or Trendy quadrants, while Economic History papers are mainly in Outlying quadrants. Figure \ref{app-fig:macro_hist_scatter} visualizes the distributions for these two fields, illustrating that while their central tendencies differ, there is substantial overlap. This overlap confirms that our framework captures within-field variation in research strategies, rather than simply acting as a proxy for subfield identity.

\begin{table}[h!]
    \centering
\begin{tabular}{lllll}
\toprule
field & Consolidating & Outlying & Trendy & Trailblazing \\
\midrule
macroeconomics & 0.31 & 0.149 & 0.474 & 0.067 \\
international & 0.257 & 0.185 & 0.42 & 0.139 \\
financial & 0.151 & 0.307 & 0.287 & 0.255 \\
economic history & 0.057 & 0.719 & 0.057 & 0.167 \\
microeconomics theory & 0.302 & 0.204 & 0.379 & 0.115 \\
industrial organization & 0.266 & 0.364 & 0.207 & 0.163 \\
econometrics \& quantitative & 0.16 & 0.276 & 0.333 & 0.23 \\
public & 0.213 & 0.342 & 0.241 & 0.203 \\
health & 0.026 & 0.31 & 0.13 & 0.535 \\
development & 0.084 & 0.37 & 0.163 & 0.383 \\
labor & 0.241 & 0.307 & 0.274 & 0.178 \\
agricultural & 0.094 & 0.57 & 0.089 & 0.247 \\
urban \& regional & 0.239 & 0.363 & 0.21 & 0.188 \\
environmental \& resource & 0.103 & 0.474 & 0.126 & 0.297 \\
behavioral & 0.098 & 0.282 & 0.24 & 0.38 \\
other & 0.079 & 0.467 & 0.083 & 0.37 \\
all & 0.207 & 0.293 & 0.293 & 0.207 \\
\bottomrule
\end{tabular}
    \caption{Distribution of novelty archetypes within economic subfields. Each row shows the fraction of papers from a given subfield that fall into each of the four novelty archetypes.}
    \label{app-tab:quadrant_stat_field}
\end{table}

\begin{figure}[h!]
    \centering
    \includegraphics[width=0.8\linewidth]{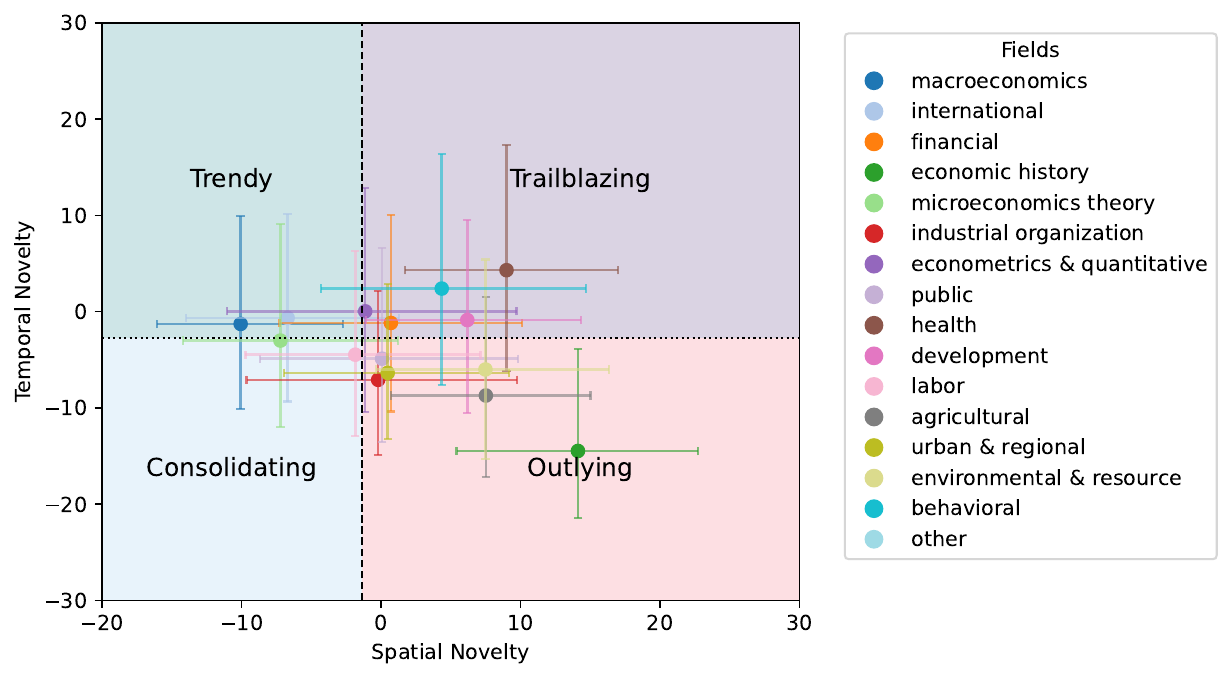}
    \caption{Median novelty profiles by economic subfield. Each point represents the median spatial and temporal novelty score for all papers in a given subfield.
    Error bars represent 25-75 inter-quartile range. 
    }
    \label{app-fig:field_novelty_sem_median_errorbar}
\end{figure}

\clearpage

\section{More Results on Typology and Impacts}
\label{app-sec:novelty_impact}
This appendix provides the detailed statistical results and robustness checks for the analysis presented in Section \ref{sec:novelty_impact}.

\subsection{Robustness of High-Citation Analysis}
\label{app-sec:high_cite_rate_quadrant}

This section provides the full results for the analysis of high-citation rates. Table \ref{app-tab:quadrant_top25_cite_rate_table} presents the detailed statistics underlying Figure \ref{fig:quadrant_top25_cite_rate}. The non-overlapping 95\% confidence intervals confirm that the differences between novelty archetypes are statistically significant.

In Table \ref{app-tab:quadrant_top25_cite_rate_table}, we first show the detailed values for Figure \ref{fig:quadrant_top25_cite_rate} in Section \ref{sec:novelty_impact_high_cite}. The high-cited rate and standard deviation are calculated from 100 resampling. We calculate the 95\% confidence interval for the values, which show that the differences between novelty types are statistically significant.

\begin{table}[h!]
    \centering
\begin{tabular}{lllll}
\toprule
Neighborhood Archetype & High-Cite Rate & SE & 95\% CI lower & 95\% CI upper \\
\midrule
Consolidating & 0.1808 & 0.00031 & 0.18023 & 0.18143 \\
Outlying & 0.1970 & 0.00027 & 0.19650 & 0.19755 \\
Trendy & 0.2914 & 0.00028 & 0.29080 & 0.29191 \\
Trailblazing & 0.3247 & 0.00032 & 0.32403 & 0.32527 \\
all & 0.2477 & 0.00001 & 0.24771 & 0.24775 \\
\bottomrule
\end{tabular}
    \caption{
    High-citation rates (Top 25\%) by novelty type.
    Statistics are calculated from 100 bootstrap iterations. A paper is defined as highly cited if its citation count is in the top 25\% of its publication-year-subfield cohort.
    }
    \label{app-tab:quadrant_top25_cite_rate_table}
\end{table}

Tables \ref{app-tab:quadrant_top10_cite_rate_table} and \ref{app-tab:quadrant_top1_cite_rate_table} test the robustness of this finding using stricter definitions of "highly cited" (top 10\% and top 1\%, respectively). The results confirm that the primary pattern holds and suggest that the relative advantage of high temporal novelty becomes even more pronounced for papers that achieve elite levels of impact. For the top 1\% threshold, high-temporal-novelty papers (in Trendy or Trailblazing Neighborhoods) are nearly twice as likely to be highly cited as their low-temporal-novelty counterparts.

\begin{table}[ht]
    \centering
\begin{tabular}{lllll}
\toprule
Neighborhood Archetype & High-Cite Rate & SE & 95\% CI lower & 95\% CI upper \\
\midrule
Consolidating & 0.0646 & 0.00021 & 0.06416 & 0.06499 \\
Outlying & 0.0736 & 0.00017 & 0.07322 & 0.07390 \\
Trendy & 0.1178 & 0.00020 & 0.11739 & 0.11815 \\
Trailblazing & 0.1345 & 0.00029 & 0.13394 & 0.13506 \\
all & 0.0973 & 0.00001 & 0.09725 & 0.09729 \\
\bottomrule
\end{tabular}
    \caption{High-citation rates (Top 10\%) by novelty type.}
    \label{app-tab:quadrant_top10_cite_rate_table}
\end{table}

\begin{table}[ht]
    \centering
\begin{tabular}{lllll}
\toprule
Neighborhood Archetype & High-Cite Rate & SE & 95\% CI lower & 95\% CI upper \\
\midrule
Consolidating & 0.0054 & 0.00006 & 0.00526 & 0.00549 \\
Outlying & 0.0074 & 0.00005 & 0.00733 & 0.00753 \\
Trendy & 0.0108 & 0.00005 & 0.01071 & 0.01091 \\
Trailblazing & 0.0137 & 0.00007 & 0.01359 & 0.01387 \\
all & 0.0093 & 0.00001 & 0.00929 & 0.00931 \\
\bottomrule
\end{tabular}
    \caption{High-citation rates (Top 1\%) by novelty type.}
    \label{app-tab:quadrant_top1_cite_rate_table}
\end{table}

\clearpage

\subsection{Analysis of Citation Count Distributions}
\label{app-sec:novelty_and_citation_counts}

Beyond the binary classification of high impact, we examine the relationship between novelty archetypes and the full distribution of citation counts. Table \ref{app-tab:quadrant_citation_count_all} reports the mean, median, and standard error (SE) for citations within each quadrant.

The results confirm the primary role of temporal novelty: papers in Trendy or Trailblazing Neighborhoods have mean citation counts more than double those of papers in Consolidating or Outlying neighborhoods. However, the analysis of median citations and the SE reveals an important secondary role for spatial novelty. Among high-temporal-novelty papers, papers in Trailblazing neighborhoods have a substantially higher median citation count than papers in Trendy neighborhoods (105.3 vs. 85.7). This pattern, combined with the lower SE for papers in Trailblazing neighborhoods, indicates that while both archetypes can produce blockbuster hits, high spatial novelty leads to a more consistently high level of impact and a less skewed citation distribution.

\begin{table}[h!]
    \centering
    \begin{tabular}{c|c|c}
    & & \\
    High Temporal          & Trendy & Trailblazing \\
                        & (Mean=269.6, Median=85.7, SE=48.54) & (Mean=270.9, Median=105.3, SE=41.40) \\
    \hline
    & & \\
    Low Temporal           & Consolidating & Outlying \\
                     & (Mean=121.4, Median=40.3, SE=26.79) & (Mean=138.0, Median=49.0, SE=30.98) \\
    \hline
    &  &  \\
         & Low Spatial & High Spatial \\
    \end{tabular}
    \caption{
    Citation distribution statistics by novelty type.
    Statistics are averaged over 100 bootstrap iterations. SE is the standard error of the mean (Std. Dev. / sqrt Mean). For the full sample, Mean=200.6, Median=65.1, CV=2.87.}
    \label{app-tab:quadrant_citation_count_all}
\end{table}

\clearpage

\subsection{Robustness of Disruption Analysis}
\label{app-sec:quadrant_disruptive_rate}

This section provides the full results for the analysis of disruptive impact. Table \ref{app-tab:quadrant_disruptive_rate_all_table} presents the statistics underlying Figure \ref{fig:quadrant_disruptive_rate_all}. The confidence intervals confirm that papers high in spatial novelty (in Outlying or Trailblazing neighborhoods) are significantly more likely to be disruptive than their low-spatial-novelty counterparts.

\begin{table}[ht]
    \centering
\begin{tabular}{lllll}
\toprule
Neighborhood Archetype & Disruption Rate & SE & 95\% CI lower & 95\% CI upper \\
\midrule
Consolidating & 0.1184 & 0.00034 & 0.11778 & 0.11910 \\
Outlying & 0.2245 & 0.00032 & 0.22385 & 0.22513 \\
Trendy & 0.1053 & 0.00024 & 0.10484 & 0.10578 \\
Trailblazing & 0.2021 & 0.00037 & 0.20135 & 0.20279 \\
all & 0.1630 & 0.00017 & 0.16264 & 0.16331 \\
\bottomrule
\end{tabular}
    \caption{Disruption rates by novelty type (full sample).}
    \label{app-tab:quadrant_disruptive_rate_all_table}
\end{table}

Tables \ref{app-tab:quadrant_disruptive_rate_top25_table} and \ref{app-tab:quadrant_disruptive_rate_top10_table} test the robustness of this finding by examining disruption rates exclusively within the subsamples of highly cited papers. Across all specifications, papers with high spatial novelty maintain a significantly higher disruption rate, confirming that this relationship is not confounded by a paper's overall level of attention.

\begin{table}[ht]
    \centering
\begin{tabular}{lllll}
\toprule
Neighborhood Archetype & Disruption Rate & SE & 95\% CI lower & 95\% CI upper \\
\midrule
Consolidating & 0.1038 & 0.00065 & 0.10254 & 0.10507 \\
Outlying & 0.2012 & 0.00073 & 0.19974 & 0.20259 \\
Trendy & 0.1062 & 0.00051 & 0.10519 & 0.10719 \\
Trailblazing & 0.1694 & 0.00052 & 0.16839 & 0.17045 \\
all & 0.1451 & 0.00029 & 0.14453 & 0.14569 \\
\bottomrule
\end{tabular}
    \caption{Disruption rates by novelty type (top 25\% cited paper only).}
    \label{app-tab:quadrant_disruptive_rate_top25_table}
\end{table}

\begin{table}[ht]
    \centering
\begin{tabular}{lllll}
\toprule
Neighborhood Archetype & Disruption Rate & SE & 95\% CI lower & 95\% CI upper \\
\midrule
Consolidating & 0.1327 & 0.00136 & 0.13000 & 0.13533 \\
Outlying & 0.2271 & 0.00122 & 0.22467 & 0.22947 \\
Trendy & 0.1219 & 0.00071 & 0.12054 & 0.12334 \\
Trailblazing & 0.1819 & 0.00082 & 0.18028 & 0.18349 \\
all & 0.1638 & 0.00048 & 0.16291 & 0.16478 \\
\bottomrule
\end{tabular}
    \caption{Disruption rates by novelty type (top 10\% cited paper only).}
    \label{app-tab:quadrant_disruptive_rate_top10_table}
\end{table}

\clearpage

\subsection{Robustness of the Advantage of Trailblazing Archetype}
\label{app-sec:quadrant_high_cite_and_disruptive_rate}
This section provides robustness checks for the advantage of papers in Trailblazing neighborhoods in achieving dual impact. Table \ref{app-tab:quadrant_disruptive_and_top25_rate_table} provides the detailed statistics for the main analysis (top 25\% cited + disruptive). Table \ref{app-tab:quadrant_disruptive_and_top10_rate_table} demonstrates that the synergistic advantage of papers in Trailblazing neighborhoods persists when using a stricter high-citation threshold (top 10\%). In both cases, papers in Trailblazing neighborhoods are significantly more likely than any other archetype to be both highly cited and disruptive.

\begin{table}[ht]
    \centering
\begin{tabular}{lllll}
\toprule
Neighborhood Archetype & Dual Impact Rate & SE & 95\% CI Lower & 95\% CI Upper \\
\midrule
Consolidating & 0.0188 & 0.00012 & 0.01854 & 0.01900 \\
Outlying & 0.0396 & 0.00016 & 0.03932 & 0.03995 \\
Trendy & 0.0309 & 0.00015 & 0.03064 & 0.03123 \\
Trailblazing & 0.0550 & 0.00017 & 0.05466 & 0.05534 \\
all & 0.0359 & 0.00007 & 0.03580 & 0.03609 \\
\bottomrule
\end{tabular}
    \caption{Probability of being both highly cited (top 25\%) and disruptive.}
    \label{app-tab:quadrant_disruptive_and_top25_rate_table}
\end{table}

We then consider papers that are both top 10\% high-cited and disruptive and show the result in in Table \ref{app-tab:quadrant_disruptive_rate_top10_table}. We can find that papers in Trailblazing neighborhoods still have advantage over other types.

\begin{table}[ht]
    \centering
\begin{tabular}{lllll}
\toprule
Neighborhood Archetype & Dual Impact Rate & SE & 95\% CI Lower & 95\% CI Upper \\
\midrule
Consolidating & 0.0086 & 0.00009 & 0.00839 & 0.00874 \\
Outlying & 0.0167 & 0.00009 & 0.01652 & 0.01688 \\
Trendy & 0.0144 & 0.00009 & 0.01419 & 0.01453 \\
Trailblazing & 0.0245 & 0.00012 & 0.02423 & 0.02469 \\
all & 0.0159 & 0.00005 & 0.01585 & 0.01603 \\
\bottomrule
\end{tabular}
    \caption{Probability of being both highly cited (top 10\%) and disruptive.}
    \label{app-tab:quadrant_disruptive_and_top10_rate_table}
\end{table}

\clearpage

\subsection{Robustness to Alternative Novelty Threshold}
\label{app-sec:split_high_75}
To ensure our findings are not sensitive to the median split used to define the novelty archetypes, we replicate the analysis using a more restrictive 75th percentile threshold. As shown in Table \ref{app-tab:quadrant_75_stat_count}, papers in Trailblazing neighborhoods becomes considerably rarer (4.0\% of the sample), underscoring the difficulty of simultaneously achieving high levels on both novelty dimensions. Despite this shift in distribution, the qualitative results remain unchanged. Tables \ref{app-tab:quadrant_75_top25_cite_rate} and \ref{app-tab:quadrant_75_disruptive_rate_all} show that temporal novelty remains the primary predictor of high citations, while spatial novelty remains the primary predictor of disruption.

\begin{table}[ht]
    \centering
    \begin{tabular}{c|c|c}
    & & \\
    High Temporal          & Trendy & Trailblazing \\
                        & (N=11086, Percent=21.0\%) & (N=2106, Percent=4.0\%) \\
    \hline
    & & \\
    Low Temporal           & Consolidating & Outlying \\
                     & (N=28488, Percent=54.0\%) & (N=11086, Percent=21.0\%) \\
    \hline
    &  &  \\
         & Low Spatial & High Spatial \\
    \end{tabular}
    \caption{Distribution of papers across the four quadrants, setting high as 75-th percentile.}
    \label{app-tab:quadrant_75_stat_count}
\end{table}

\begin{table}[ht]
    \centering
\begin{tabular}{lllll}
\toprule
Neighborhood Archetype & High-Cite Rate & SE & 95\% CI lower & 95\% CI upper \\
\midrule
Consolidating & 0.2136 & 0.00018 & 0.21322 & 0.21395 \\
Outlying & 0.2188 & 0.00038 & 0.21809 & 0.21957 \\
Trendy & 0.3401 & 0.00039 & 0.33932 & 0.34085 \\
Trailblazing & 0.3756 & 0.00097 & 0.37365 & 0.37746 \\
all & 0.2477 & 0.00001 & 0.24771 & 0.24775 \\
\bottomrule
\end{tabular}
    \caption{High-Citation rates (Top 25\%) using 75th percentile threshold.}
    \label{app-tab:quadrant_75_top25_cite_rate}
\end{table}

We also check the disruptive rate using this typology, and show the result in Table \ref{app-tab:quadrant_75_disruptive_rate_all}. We can see that types with high spatial novelty (in Outlying or Trailblazing neighborhoods) have much higher high-cited rate compared to types with low spatial novelty.

\begin{table}[ht]
    \centering
\begin{tabular}{lllll}
\toprule
Neighborhood Archetype & Disruption Rate & SE & 95\% CI lower & 95\% CI upper \\
\midrule
Consolidating & 0.1442 & 0.00024 & 0.14370 & 0.14463 \\
Outlying & 0.2508 & 0.00040 & 0.25004 & 0.25161 \\
Trendy & 0.1150 & 0.00031 & 0.11441 & 0.11561 \\
Trailblazing & 0.2075 & 0.00091 & 0.20576 & 0.20933 \\
all & 0.1630 & 0.00017 & 0.16264 & 0.16331 \\
\bottomrule
\end{tabular}
    \caption{Disruption rate using 75th percentile threshold (full sample).}
    \label{app-tab:quadrant_75_disruptive_rate_all}
\end{table}

\clearpage

\subsection{Field-Level Analysis}
\label{app-sec:field_level}
To ensure our findings are not driven by compositional effects across disciplines, we replicate the analysis by constructing the novelty scores and examining impact rates within each subfield. Tables \ref{app-tab:high_cite_by_field} and \ref{app-tab:disruptive_rate_by_field} present the results. Despite variation in the magnitude of effects, the core qualitative patterns hold across nearly all subfields: papers high in temporal novelty (in Trendy or Trailblazing neighborhoods) consistently have higher citation rates, while papers high in spatial novelty (in Outlying or Trailblazing neighborhoods) consistently have higher disruption rates. This confirms that our framework captures general pathways to scientific influence.

\begin{table}[h!]
    \centering
    \small
\begin{tabular}{lllll}
\toprule
field & Consolidating & Outlying & Trendy & Trailblazing \\
\midrule
macroeconomics & 0.196 & 0.194 & 0.291 & 0.322 \\
 & (0.0012) & (0.0009) & (0.0009) & (0.0013) \\
international & 0.149 & 0.175 & 0.324 & 0.343 \\
 & (0.0012) & (0.0010) & (0.0013) & (0.0014) \\
financial & 0.179 & 0.195 & 0.317 & 0.293 \\
 & (0.0013) & (0.0011) & (0.0010) & (0.0016) \\
economic history & 0.178 & 0.184 & 0.319 & 0.273 \\
 & (0.0029) & (0.0018) & (0.0017) & (0.0032) \\
microeconomics theory & 0.186 & 0.227 & 0.238 & 0.362 \\
 & (0.0009) & (0.0007) & (0.0007) & (0.0011) \\
industrial organization & 0.162 & 0.227 & 0.248 & 0.361 \\
 & (0.0012) & (0.0011) & (0.0010) & (0.0014) \\
econometrics \& quantitative & 0.160 & 0.197 & 0.286 & 0.361 \\
 & (0.0011) & (0.0008) & (0.0007) & (0.0012) \\
public & 0.187 & 0.196 & 0.278 & 0.339 \\
 & (0.0012) & (0.0010) & (0.0010) & (0.0015) \\
health & 0.218 & 0.161 & 0.319 & 0.289 \\
 & (0.0019) & (0.0016) & (0.0016) & (0.0019) \\
development & 0.212 & 0.175 & 0.312 & 0.291 \\
 & (0.0015) & (0.0013) & (0.0014) & (0.0016) \\
labor & 0.183 & 0.177 & 0.308 & 0.333 \\
 & (0.0011) & (0.0008) & (0.0010) & (0.0013) \\
agricultural & 0.191 & 0.169 & 0.322 & 0.283 \\
 & (0.0027) & (0.0019) & (0.0021) & (0.0035) \\
urban \& regional & 0.233 & 0.152 & 0.312 & 0.275 \\
 & (0.0027) & (0.0022) & (0.0026) & (0.0029) \\
environmental \& resource & 0.163 & 0.208 & 0.283 & 0.327 \\
 & (0.0020) & (0.0019) & (0.0018) & (0.0023) \\
behavioral & 0.185 & 0.222 & 0.272 & 0.309 \\
 & (0.0014) & (0.0013) & (0.0015) & (0.0020) \\
other & 0.136 & 0.186 & 0.323 & 0.328 \\
 & (0.0024) & (0.0020) & (0.0025) & (0.0027) \\
\bottomrule
\end{tabular}
    \caption{High-citation rates (top 25\%) by novelty archetype, by subfield. Table entries report the mean high-citation rate with bootstrapped standard errors in parentheses. Novelty archetypes are defined within each subfield.}
    \label{app-tab:high_cite_by_field}
\end{table}

\begin{table}[h!]
    \centering
    \small
\begin{tabular}{lllll}
\toprule
field & Consolidating & Outlying & Trendy & Trailblazing \\
\midrule
macroeconomics & 0.104 & 0.192 & 0.081 & 0.158 \\
 & (0.0010) & (0.0010) & (0.0006) & (0.0013) \\
international & 0.137 & 0.239 & 0.109 & 0.196 \\
 & (0.0011) & (0.0012) & (0.0009) & (0.0013) \\
financial & 0.073 & 0.165 & 0.075 & 0.121 \\
 & (0.0010) & (0.0011) & (0.0008) & (0.0013) \\
economic history & 0.181 & 0.231 & 0.210 & 0.139 \\
 & (0.0032) & (0.0023) & (0.0023) & (0.0031) \\
microeconomics theory & 0.052 & 0.135 & 0.055 & 0.105 \\
 & (0.0006) & (0.0007) & (0.0004) & (0.0008) \\
industrial organization & 0.100 & 0.203 & 0.139 & 0.219 \\
 & (0.0009) & (0.0015) & (0.0011) & (0.0013) \\
econometrics \& quantitative & 0.105 & 0.186 & 0.068 & 0.158 \\
 & (0.0009) & (0.0009) & (0.0006) & (0.0011) \\
public & 0.151 & 0.244 & 0.127 & 0.270 \\
 & (0.0011) & (0.0014) & (0.0009) & (0.0015) \\
health & 0.162 & 0.336 & 0.151 & 0.268 \\
 & (0.0018) & (0.0024) & (0.0015) & (0.0020) \\
development & 0.253 & 0.376 & 0.262 & 0.332 \\
 & (0.0019) & (0.0019) & (0.0020) & (0.0029) \\
labor & 0.144 & 0.257 & 0.198 & 0.234 \\
 & (0.0010) & (0.0011) & (0.0011) & (0.0014) \\
agricultural & 0.353 & 0.308 & 0.379 & 0.367 \\
 & (0.0033) & (0.0028) & (0.0033) & (0.0041) \\
urban \& regional & 0.200 & 0.269 & 0.107 & 0.203 \\
 & (0.0028) & (0.0030) & (0.0019) & (0.0028) \\
environmental \& resource & 0.159 & 0.282 & 0.221 & 0.346 \\
 & (0.0019) & (0.0021) & (0.0021) & (0.0029) \\
behavioral & 0.049 & 0.143 & 0.058 & 0.104 \\
 & (0.0009) & (0.0013) & (0.0009) & (0.0013) \\
other & 0.313 & 0.269 & 0.163 & 0.288 \\
 & (0.0029) & (0.0025) & (0.0022) & (0.0024) \\
\bottomrule
\end{tabular}
    \caption{Disruption rates by novelty archetype, by subfield. Table entries report the mean disruption rate with bootstrapped standard errors in parentheses. Novelty archetypes are defined within each subfield.}
    \label{app-tab:disruptive_rate_by_field}
\end{table}

\clearpage

\section{Prompts}
\label{app-sec:prompt}
\begin{lstlisting}[breaklines,caption={Prompt for comprehensive summary.}]
"""
Generate an in-depth 1000-2000 word structured summary of this economics research paper that simultaneously reconstructs its content and provides critical analysis. Adhere to these specifications:
### 1. Extended Structural Reconstruction (800-1200 words)
For each section, include:
#### Introduction  
- Research context & motivations (2-3 paragraphs)  
- Explicitly stated hypotheses/research questions  
- Literature review synthesis: Map key citations to research gaps  
#### Theoretical Framework  
- Foundational Theories: Outline the core economic theories and models that underpin the research.  
- Assumptions and Constructs: Detail the theoretical assumptions made and define key constructs.  
- Conceptual Integration: Explain how these theories inform the research questions, methodology, and analysis.  
- Critical Perspective: Evaluate the strengths and limitations of the theoretical approach in addressing the identified research gaps.
#### Methodology  
- Data sources: Providers, timeframes, inclusion/exclusion criteria  
- Analytical framework: Equations/diagrams with plain-English explanations  
- Technical procedures: Step-by-step workflow  
#### Experiments/Analysis  
- Implementation chronology  
- Control variables/benchmarks  
- Sensitivity testing documentation  
#### Results  
- Primary/secondary findings with statistical significance  
- Non-obvious patterns in data  
- Failed hypotheses (if any)  
#### Discussion  
- Mechanisms explaining results  
- Comparative analysis with 3-5 key cited works  
- Policy recommendations with implementation pathways  

### 2. Comprehensive Evaluation (500-800 words)
Develop dedicated critique sections:
#### Innovation Audit  
- Novelty: Conceptual | Methodological | Empirical  
- Patentable prior art comparison  
- Disruptiveness score (1-5) with rationale  
#### Methodological Rigor  
- STROBE/RDD checklist compliance  
- Data limitations  
- Counterfactual analysis completeness  
#### Communication Effectiveness  
- Readability metrics: Flesch-Kincaid grade level | Jargon density  
- Argumentation coherence mapping  
- Visual aid effectiveness assessment  
#### Scalability Potential  
- Replication constraints analysis  
- Generalizability (Contexts/Data/Methods)  
- Cost-benefit simulation  

### 3. Enhancement Protocol (200-300 words)
Provide actionable improvements for:  
- Theory-building extensions  
- Methodological upgrades  
- Policy translation frameworks  
### Format Requirements:
- Use MLA citation style for all references  
- Include 4-6 verbatim quotes exemplifying key strengths/weaknesses  
- Present technical content through Definition Boxes  
- Highlight critical insights in **bold**  
- Maintain a graduate-level academic tone  
- Deliver output as a hybrid document integrating summary and peer-review elements  

Here is the paper:
{Paper_Content}
"""
\end{lstlisting}

\begin{lstlisting}[breaklines,caption={Prompt for Paragraph insight.}]
"""
You are an expert in economics with experience in reading and summarizing academic research papers. 
I have a detailed summary of an economics research paper, and I would like you to write a shorter, more concise summary while retaining all the key information.

**Goal:** 
Write a summary of several paragraphs that clearly conveys the main contributions, innovations, and findings of the paper. 
The summary should be clear and informative enough for a researcher to understand the core of the paper without reading it.

**Instructions:**  
- Keep the summary concise but comprehensive.  
- Clearly explain the research question, methodology, data, and findings.  
- Highlight the paper's main contributions and innovations.  
- Avoid unnecessary details or lengthy explanations.  
- Maintain a formal and academic tone.  

Here is the detailed summary of the paper:
{paper_content}

Please write the shorter summary below:
"""
\end{lstlisting}

\begin{lstlisting}[breaklines,caption={Prompt for Topic insight.}]
"""
Please extract the research topic of the given text in one sentence strictly following this format:  
"The economic subfield is [subfield name], the main research subject is [subject], it focuses on [specific issue], and it uses [approach]."
Your response should be at most 40 words.  

Here is the input text: 
{paper_content}

Please write the response below:
"""
\end{lstlisting}

\begin{lstlisting}[breaklines,caption={Prompt for Question-Conclusion insight.},literate={≤}{{$\leq$}}1]
"""
Please summarize the research question and conclusions of the given text according to the following structure (≤100 words):  

1. Core Research Question: [What the paper seeks to answer, framed as a clear question]  
2. Approach and Hypothesis: [For empirical studies, describe the hypothesis; for theoretical papers, state the key assumption]  
3. Key Findings: [Discoveries, including a comparison with expectations or existing literature]  
4. Novelty and Implications: [What makes this conclusion different or impactful]

Here is the input text: 
{paper_content}

Please write the summary below:
"""
\end{lstlisting}

\begin{lstlisting}[breaklines,caption={Prompt for Methodology insight.},literate={≤}{{$\leq$}}1 {β}{{$\beta$}}1 {ε}{{$\epsilon$}}1]
"""
Please summarize the methodology of the given text in a structured format (≤80 words):  

- Data Source: [Dataset name, time span, sample size]  
- Core Model: [Mathematical expression, e.g., Y = βX + ε]  
- Analytical Techniques: [e.g., IV estimation, text mining process]  
- Assumptions: [Key economic assumptions, if applicable]  
- Methodological Innovation: [How this method improves upon or differs from existing methods]  

Here is the input text: 
{paper_content}

Please write the summary below:
"""
\end{lstlisting}

\begin{lstlisting}[breaklines,caption={Prompt for Theory insight.},literate={≤}{{$\leq$}}1]
"""
Please summarize the theoretical contributions of the given text according to the following structure (≤100 words):  

1. Fundamental Theory: [The main theoretical framework it builds upon]  
2. Extensions: [Modified assumptions, newly introduced mechanisms]  
3. Theoretical Advances: [Innovations in key formulas or theorem proofs]  
4. Scope of Explanation: [Extensions of the new theory to applicable scenarios]  
5. Empirical Relevance: [How the theoretical insights connect with empirical findings, if applicable]  

Here is the input text: 
{paper_content}

Please write the summary below:
"""
\end{lstlisting}

\end{document}